\documentclass[fleqn,usenatbib]{mnras}

\usepackage{newtxtext,newtxmath}

\usepackage[T1]{fontenc}

\DeclareRobustCommand{\VAN}[3]{#2}
\let\VANthebibliography\thebibliography
\def\thebibliography{\DeclareRobustCommand{\VAN}[3]{##3}\VANthebibliography}

\usepackage{graphicx}	
\usepackage{amsmath}	
\usepackage{bm}
\usepackage{mathtools}

\title[linear operator theory]{Linear operator theory of phase mixing}

\author[K. Darling and L. M. Widrow]{
Keir Darling \thanks{E-mail: keir.darling@queensu.ca}
and 
Lawrence M. Widrow
\\
Department of Physics, Engineering Physics \& Astronomy, Queen's University, Stirling Hall, Kingston ON K7L 3N6, Canada\\
}

\date{Accepted XXX. Received YYY; in original form ZZZ}

\pubyear{2024}

\begin{document}
\label{firstpage}
\pagerange{\pageref{firstpage}--\pageref{lastpage}}
\maketitle

\begin{abstract}
	
We study solutions of the collisionless Boltzmann equation (CBE) in a functional Koopman representation. This facilitates the use of linear spectral techniques characteristic of the analysis of Schr{\"o}dinger-type equations. For illustrative purposes, we consider the classical phase mixing of a non-interacting distribution function in a quartic potential. Solutions are determined perturbatively relative to a harmonic oscillator. We impose a form of coarse-graining by choosing a finite dimensional basis to represent the distribution function and time evolution operators, which sets a minimum length scale on phase space structure. We observe a relationship between the dimension of the representation and the multiplicity of the harmonic oscillator eigenvalues. System dynamics are understood in terms of degenerate subspaces of the linear operator spectra. Each subspace is associated with a mode of the harmonic oscillator, the first two being bending and breathing structures. The quartic potential splits the degenerate eigenvalues within each subspace. This facilitates the formation of spiral structure as deformations from the harmonic oscillator modes. We ultimately argue that this construction provides a promising avenue for study of self-interacting systems experiencing phase mixing, which is an outstanding problem in the context of the Gaia DR2 vertical phase space spirals.

\end{abstract}

\begin{keywords}
Galaxy: disc -- Galaxy: kinematics and dynamics -- Galaxy: structure.
\end{keywords}



\section{Introduction}\label{section:introduction}

In this paper we investigate the relaxation of collisionless systems through phase mixing. In a statistical description, the macrostate of a system is specified by the phase space distribution function, $f$. This quantifies the probability that a particle exists within an infinitesimal volume of phase space \citep{sethna2006}. Liouville's theorem requires that $f$ is conserved along orbits, and it therefore satisfies the collisionless Boltzmann equation (CBE) \citep{arnold1989}. For $s$ spatial degrees of freedom, this is equivalent to the incompressible flow of a $2s$ dimensional fluid in a velocity field specified by Hamilton's equations. An introductory description of this can be found in \cite{binneytremain2008}, but we summarize as follows. Let us assume an anharmonic potential in which orbital frequency of test particles is dependent on their amplitudes of oscillation. In the fluid analogy, this means that vorticity of the velocity field depends on the spatial coordinate. A distribution out of equilibrium with such a potential will deform with time as packets of density with different energies orbit at varying frequencies. For a fixed conservative Hamiltonian this continues indefinitely, with different energy orbits becoming increasingly out of phase with each other. In this process, the scale of structure in the distribution decreases. Eventually when the scale becomes so small that adjacent wraps of the mixed distribution become indistinguishable, the system has equilibrated. 

Phase mixing in general leads to complex structures, especially for the $s=3$ case present in galactic dynamics and cosmology \citep{tremaine1999,stiff2003,abel2012}. Even for $s=1$, which applies to considerations of the vertical motion in the Galactic disc, this process is not trivial. With astrometric data from \cite{gaia2018}, a one armed spiral was observed in the vertical phase space of Solar Neighborhood stars \citep{antoja2018}. In reality this is a $s=3$ system, as it is unlikely that the vertical dynamics are decoupled from motion in the plane \citep{hunt2021}, but much attention has been given to the structure in the univariate case \citep{binney2018,darling2018,bennet2018,bennet2021}. At present, no models have reproduced the exact form of the Gaia spiral. 

In \cite{darling2019}, it was suggested that the phase mixing process can be represented with the discrete spectrum of a linear time-evolution operator of finite dimension. There, eigenfunctions were estimated numerically from the full temporal history of a system by applying Dynamic Mode Decomposition (DMD) \citep{mezic2005,rowleyMezic2009,kutz2016} to $N$-body simulations. This served to investigate the claim that self-gravity should not be ignored in phase mixing \citep{darling2018}, as well as to explore the representation of this process with persistent oscillatory structures. Stable oscillations such as bending and breathing modes were observed in a self-interacting system in an anharmonic potential. When modifying the relative dominance of self-interaction and anharmonic forcing, these oscillatory structures were more prominent the closer the system was to purely self-interacting. For stronger anharmonic forcing, they were deformed to include spiral structure. 

DMD is closely related to Koopman theory \citep{koopman1931}, which supposes that a complex, potentially nonlinear system can be represented as a simpler linear one by studying its evolution in terms of observable functions of its state space. This concept was used to interpret the results in \cite{darling2019}, arguing that the binning of $N$-body simulations constituted a mapping to observables. Because of the numerical nature of that work, it was difficult to study the supposed mechanism of phase mixing from discrete modes, or establish a concrete connection to Koopman theory. The DMD approach also draws comparison to the use of multichannel Singular Spectrum Analysis (mSSA) \citep{msaa2021,msaa2023}. In both cases, principle component analysis (PCA) based techniques are applied to time-series data. A novel aspect of the mSSA work was the temporal analysis of basis function expansion coefficients. This motivates treating the expansion coefficients as Koopman observables, rather than the distribution function.

In the present work, we aim to investigate a mechanism of phase mixing with a discrete linear operator spectrum, emphasizing the role of representation scale, and properties of the spectrum. The essential premise is that the minimum length scale and degeneracy of evolution operator eigenvalues depend on the dimension of the representation. The degenerate eigenvalues define subspaces of the operator spectrum. Splitting of these eigenvalues by anharmonic forcing causes differential rotation among the subspace eigenvectors, which manifests as phase mixing in $f$. System dynamics are determined in terms of basis function expansion coefficients, which we treat as functionals of $f$, adopting a functional Koopman formalism for our calculations. In doing so, we apply techniques from degenerate perturbation theory of the Schr{\"odinger} equation. All of our calculations are carried out symbolically.

This paper is organized as follows. In Section \ref{section:setup} we define our coordinates and  Hamiltonian, as well as the vector spaces in which we perform our calculations. In section \ref{section:dynamics} we define the time evolution operator for functionals of $f$, and then restate the problem in this formalism. In Section \ref{section:matrix} we define a matrix representation for our operators, and compute the harmonic oscillator spectrum. Section \ref{section:anharmonic_potential} contains the perturbative treatment of the anharmonic potential, and a description of the eigenvalue splitting mechanism for phase mixing. Sections \ref{section:discussion} and \ref{section:conclusion} contain a discussion of connections to other works, and concluding remarks on foreseeable extensions. 

\section{Preliminaries}\label{section:setup}

Consider a two-dimensional phase space comprising position $\bar{q}$ and momentum $ \bar{p} $. We concern ourselves with the phase space distribution function, $f(\bar{q},\bar{p})$. This is defined such that, 

\begin{equation}\label{eq:df_definition}
	\mathrm{P}\bigl(\bar{q},\bar{p}\in\mathcal{R}\bigr) = \int_\mathcal{R} d\bar{q}d\bar{p} \  f(\bar{q},\bar{p}),
\end{equation}

\noindent is the probability of finding a particle in the region of phase space, $\mathcal{R}$.  The dynamics of $f$ are prescribed by the Hamiltonian density, $h(\bar{q},\bar{p})$, according to the CBE,

\begin{equation}\label{eq:CBE}
	\frac{\partial f}{\partial t} + [f,h] = 0.
\end{equation}

\noindent Here $ [f , h ] = \frac{\partial f}{\partial q}\frac{\partial h}{\partial p} - \frac{\partial f}{\partial p}\frac{\partial h}{\partial q} $ denotes the Poisson bracket.

\subsection{System Hamiltonian}

We begin with the Hamiltonian density for a harmonic oscillator with frequency $\kappa$,

\begin{equation}\label{key}
	h_0(\bar{q},\bar{p}) = \tfrac{1}{2}\left(\bar{p}^2 + \kappa^2 \bar{q}^2\right). 
\end{equation}

\noindent This is to be understood as the standard kinematic term plus the leading term in a Taylor series of any even potential. By Jeans' theorem, the Hamiltonian defines the coordinate dependence of the equilibrium distribution function. Let us choose

\begin{equation}\label{key}
	f_0(\bar{q},\bar{p}) = \frac{1}{Z}\mathrm{e}^{-\frac{1}{2}\beta h_0(\bar{q},\bar{p})}.
\end{equation}

\noindent Here, $\beta$ is a coldness parameter inversely proportional to the velocity dispersion, and $Z$ is the partition function. It is convenient to work with scaled dimensionless coordinates, so we let

\begin{equation}\label{eq:coordinate_transformation}
	\bar{q} = \tfrac{1}{\kappa}\sqrt{\tfrac{2}{\beta}}q, \ \ \bar{p} = \sqrt{\tfrac{2}{\beta}}p.
\end{equation}

\noindent In these coordinates, the Hamiltonian density and distribution function become

\begin{equation}\label{eq:scaled_harmonic}
	h_0(q,p) = \frac{1}{\beta} \left(q^2 + p^2\right), \ \ f_0(q,p) = \frac{1}{Z}\mathrm{e}^{-\frac{1}{2}(q^2+p^2)}.
\end{equation}

Now we add an anharmonic correction of the form $\varphi\bar{q}^4$, representative of the next term in the even Taylor series. Our total Hamiltonian density in scaled coordinates is

\begin{equation}\label{eq:full_hamiltonian}
	h(q,p) = h_0(q,p) + \varphi \frac{4}{\kappa^4\beta^2}q^4.
\end{equation}

\noindent The factor $4\kappa^{-4}\beta^{-2}$ comes from the coordinate transformation in equation \ref{eq:coordinate_transformation}. We will use $\varphi = -\frac{\beta}{12}(\frac{\kappa}{2})^4$ in our calculations. This is chosen to assure stable orbits for a range of initial conditions while introducing vorticity that decreases with position, of which the effects can be seen within a few dynamical times.  

\subsection{Space of bivariate functions}\label{section:function_space}

Any configuration of $f$ is an element of a function space defined for the independent variables $q$ and $p$. Denote by $\mathscr{H}$ the space of functions of $q$ and $p$ on a domain $\mathcal{D}$. We treat $\mathscr{H}$ as a Hilbert space equipped with an inner product, which we denote $\langle g_1,g_2\rangle$ for any $g_1,g_2\in\mathscr{H}$ (see equation \ref{eq:inner_product}). We further assume that there exists a set of linearly independent functions $\{e_{j,k}\}$ that form a basis in $\mathscr{H}$. Any function $g\in\mathscr{H}$ can be written in terms of the basis functions by projecting onto them (equation \ref{eq:function_space_expansion}).

\subsection{Conjugate space of functionals}

Given the function space $\mathscr{H}$, one may consider another, distinct vector space housing its functionals. Such a space is called \textit{dual} to $\mathscr{H}$, and is denoted $\mathscr{H}^*$ (also a Hilbert space). For the purposes of this work, we refer to a functional of $f$ as any linear mapping 

\begin{equation}\label{eq:functional_definition}
	G[f] = \langle f,g\rangle, 
\end{equation}

\noindent which takes the input function, here the distribution function $f$, and maps it to $\mathbb{C}$ (the complex numbers) by integration with a given $g$. For every $g\in\mathscr{H}$, there is a unique $G\in\mathscr{H}^*$ given by equation \ref{eq:functional_definition} (Riesz representation theorem, see for example \cite{conway1994}). In the present context, the input function will always be $f$, and it is treated as a variable relative to a functional $G[f]\in\mathscr{H}^*$ in the same sense that $q$ and $p$ are variables with respect to a function $g(q,p)\in\mathscr{H}$. The functions $f,g\in\mathscr{H}$ and the functional $G\in\mathscr{H}^*$ are further related by the functional derivative, $ \frac{\delta G}{\delta f} = g(q,p)$.

The dual space $\mathscr{H}^*$ has an inner product and basis that can be understood in terms of the corresponding constructions in $\mathscr{H}$. Given a basis $\{e_{j,k}\}$ of $\mathscr{H}$, there exists a basis for $\mathscr{H}^*$,  which we denote $\{E_{j,k}\}$. The two are related by the biorthonormal condition, $E_{j,k}[e_{j',k'}] = \delta_j^{j'}\delta_k^{k'}$. The inner product of any two $G_1,G_2\in\mathscr{H}^*$ is denoted $\langle G_1,G_2\rangle_*$, and defined in equation \ref{eq:dual_inner_product}. The subscript asterisk signifies that the operation is in $\mathscr{H}^*$.

Finally we highlight some important quantities that exist in $\mathscr{H}^*$. The total energy, which is the expectation value of the Hamiltonian with respect to $f$, is the functional $H[f] = \langle f,h \rangle$. Additionally, if we project $f$ onto a set of basis functions in $\mathscr{H}$ as in equation \ref{eq:function_space_expansion}, the coefficients are the functionals $E_{j,k}^\dagger$. The $^\dagger$ operation denotes complex conjugation on scalars, and the conjugate transpose on vectors and matrices. If we compute the dynamics of $E_{j,k}$ according to $H$ in $\mathscr{H}^*$, we obtain the time-evolution of the basis function expansion coefficients for $f$ in $\mathscr{H}$.

\subsection{Particular choice of basis functions}\label{section:choice_of_basis}

We choose as a basis for $\mathscr{H}$ products of univariate Gaussian-Hermite functions. That is,

\begin{equation}\label{key}
\begin{aligned}
	e_{j,k}(q,p) = N_{j,k} f_0(q,p)P_j(q)P_k(p),
\end{aligned}
\end{equation}

\noindent where $ P_j(x) $ denotes the $ j $th Hermite polynomial of either $q$ or $p$, and $N_{j,k}$ is a normalization constant. The corresponding functionals are denoted $E_{j,k}[f] = \langle f,e_{j,k}\rangle$. In Appendix \ref{A:definitions} we define the polynomials $P_j$, the coefficients $N_{j,k}$, and state two recurrence relations for the Hermite polynomials we will make use of in Sections \ref{section:A_0_elements} and \ref{section:A_4_elements}.

\section{Dynamics}\label{section:dynamics}

In what follows, we will use the notion of a flow map. This is an operator on $\mathscr{H}$ denoted $\hat{S}^t$, which takes an initial state $f$ to another state at time $t$, $\hat{S}^tf$. The particular action of $\hat{S}^t$ is prescribed by the CBE (equation \ref{eq:CBE}).

\subsection{Time evolution operator for functionals}\label{section:functional_dynamics}

Knowing the flow map $\hat{S}^t$ is tantamount to solving equation \ref{eq:CBE}. Here, we leverage \cite{koopman1931} to effectively apply the flow map, without dealing explicitly with the CBE. Discussion of the Koopman formalism in the context of astrophysics can be found in \cite{darling2019} and \cite{darling2021}, but for a more rigorous description of the functional realization we will use here, see \cite{mezic2020}. Briefly, the idea is that rather than consider the distribution function directly, one can instead study the dynamics of ``observables''. In the case of a partial differential equation like the CBE, the observables take the form of functionals, $G[f]$. Let us define a linear time evolution operator that acts on $\mathscr{H}^*$, which we denote $ \hat{U}^t $. This is called the Koopman operator, and its action on any $G\in\mathscr{H}^*$ is,

\begin{equation}\label{eq:U_definition}
	\hat{U}^tG[f] = G[\hat{S}^tf].
\end{equation}

\noindent That is, $\hat{U}^t$ applies the flow map to the argument function $f$, but does not change the form of the functional $G[f]$. If it is easier to determine the action of $\hat{U}^t$ than $\hat{S}^t$, one can compute the functional $\hat{U}^tG[f]$, which encodes information about $f$ at time $t$. Subsequently, if we know how to infer $f$ from a functional or set of functionals, we can compute the future state $\hat{S}^tf$. 

Treating $\hat{U}^tG$ as a function of $t$, we may write its infinitesimal change with respect to $t$ as

\begin{equation}\label{eq:generator_definition_1}
	\frac{d}{dt}\hat{U}^tG[f]  =\lim_{\tau\rightarrow 0} \frac{  \hat{U}^{t+\tau} G[f]-\hat{U}^tG[f]}{\tau}.
\end{equation} 

\noindent To evaluate the $\tau$ limit, we separate out the $\hat{U}^t$ (by linearity), and apply equation \ref{eq:U_definition} for $\hat{U}^\tau G$. That is,

\begin{equation}\label{eq:generator_definition_2}
	\frac{d}{dt}\hat{U}^tG[f]  =\hat{U}^t\left(\lim_{\tau\rightarrow 0} \frac{  G[\hat{S}^\tau f]-G[f]}{\tau}\right).
\end{equation} 

\noindent For small $\tau$, $\hat{S}^\tau f = f + \tau [h,f] + \mathcal{O}(\tau^2)$. Replacing $G[\hat{S}^\tau f]$ in equation \ref{eq:generator_definition_2} with this takes care of the limit, and we are left with 

\begin{equation}\label{eq:koopman_ode}
\frac{d}{dt}\hat{U}^tG[f]  = \Bigl\langle [h,f], \frac{\delta }{\delta f} \hat{U}^tG\Bigr\rangle \coloneqq \hat{A}\Bigl(\hat{U}^tG[f]\Bigr) .
\end{equation}

\noindent Here we have defined as $\hat{A}$ the infinitesimal generator of $\hat{U}^t$. This is another operator on $\mathscr{H}^*$, and will be the focus of much of this work. To be clear, the action of $\hat{A}$ on any $G\in\mathscr{H}^*$ is 

\begin{equation}\label{eq:generator_definition}
	\hat{A}G = \Bigl\langle [h,f], \frac{\delta G}{\delta f}\Bigr\rangle = \Bigl\langle f, \left[\frac{\delta G}{\delta f},\frac{\delta H}{\delta f}\right]\Bigr\rangle.
\end{equation}

\noindent The final equality is obtained by expanding the Poisson bracket and performing integration by parts with respect to $p$ on the first term and $q$ on the second. This final form is called a Morrison bracket, which originates in plasma physics \citep{morrison1980}. We denote this multiplicative operation between vectors in $\mathscr{H}^*$ by $[G,H]_f$, indicating that it is the bracket between $G$ and $H$ with respect to $f$. We prefer this form, as it makes clear that a functional $G$ acted on by $\hat{A}$ is another vector in $\mathscr{H}^*$. For $g=\frac{\delta G}{\delta f}$, $\hat{A}G$ maps the expectation value of $g$ to that of $[g,h]$. An operator of this form has appeared in the gravitational context previously in for example \cite{perez2005}, where the Morrison bracket is used to form a so-called functional Vlasov equation. 

Equation \ref{eq:koopman_ode} is a Schr{\"o}dinger-type equation. It follows that the operator $\hat{U}^t$ satisfies an ordinary differential equation, and has the general solution

\begin{equation}\label{eq:koopman_operator}
	\hat{U}^t  = \mathrm{e}^{\int_0^t d\tau \hat{A}}.
\end{equation}

\noindent The reader familiar with quantum mechanics (QM) may find that this resembles the Heisenberg representation, with $\hat{A}$ analogous to the Hamiltonian operator. Like in QM, we can construct solutions to equation \ref{eq:koopman_ode} by solving the associated eigenvalue problem of $\hat{A}$.

\subsection{Time evolution of $f$}

Denote the $j$th eigenvalue and eigenfunctional of $\hat{A}$ by $\lambda_j$ and $\Psi_j[f] $ respectively. The eigenvalue problem is then, $\hat{A}\Psi_j=\lambda_j\Psi_j$. Let $\psi_j(q,p) = \frac{\delta \Psi_j}{\delta f}$. For convenience, we call this quantity an eigenfunction of $\hat{A}$, but really it is an eigenfunction of the Liouville operator, $[h, \cdot]$. Suppose that $\{\psi_j\}$ forms a basis of $\mathscr{H}$, and $\{\Psi_j\}$ of $\mathscr{H}^*$.

In this paper, we focus on the case where $\frac{\partial \hat{A}}{\partial t}=0$, which corresponds to a time-independent potential. This holds for our target Hamiltonian in equation \ref{eq:full_hamiltonian}, but is not true in general for a self-consistent distribution function satisfying both the CBE and Poisson equation. To study the response to both an external potential and self-consistent density perturbations, one must consider $\frac{\partial \hat{A}}{\partial t}\neq0$. We will discuss this briefly in Sections \ref{section:discussion} and \ref{section:conclusion}, but treatment of that problem is not in the scope of this work. 

In the time-independent generator case, equation \ref{eq:koopman_operator} simplifies to $\hat{U}^t=\mathrm{e}^{\hat{A}t}$, and we have

\begin{equation}
	\hat{U}^t\Psi_j[f] = \Psi_j[\hat{S}^tf] = \mathrm{e}^{\lambda_j t}\Psi_j[f].
\end{equation}

\noindent By conjugate symmetry of the inner product, the expansion coefficient of $f$ onto $\psi_j$ at time $t$ is then

\begin{equation}
\langle \psi_j,\hat{S}^tf\rangle = \Psi^\dagger_j[\hat{S}^tf] = \mathrm{e}^{\lambda_j^\dagger t }\langle \psi_j,f\rangle.
\end{equation}

\noindent Applying $\hat{S}^t$ to equation \ref{eq:function_space_expansion} and using this result,  $\hat{S}^tf$ is 

\begin{equation}\label{eq:f_expansion_t}
	\hat{S}^t f(q,p) = \sum_{j=1}^\infty\langle \psi_j, f\rangle\mathrm{e}^{\lambda_j^\dagger t}\psi_j(q,p).
\end{equation}

\subsection{Restatement of the problem}

Now that we have established the formalism in which we will carry out our calculations, it is advantageous to restate the original problem. Recall that the target Hamiltonian is stated in equation \ref{eq:full_hamiltonian}. In terms of total energy functionals, this corresponds to $H = H_0 + \varphi H_4$, where $H_0 = \langle f,h_0\rangle $ and $H_4 = \langle f, \frac{4}{\kappa^4\beta^2} q^4\rangle $. It follows that the generators are $	\hat{A}_0 = [\cdot,H_0]_f$, $\hat{A}_4 = [\cdot,H_4]_f$, and $\hat{A} = \hat{A}_0 + \varphi\hat{A}_4$. Our goal then is to compute the spectrum of $\hat{A}$, so that we can use equation \ref{eq:f_expansion_t}. 

\section{Matrix Representation}\label{section:matrix}

To carry out calculations, we map the quantities in $\mathscr{H}^*$ to finite dimensional matrices and vectors. We begin by choosing a finite dimension, $M\in\mathbb{N}$. If we take all bivariate Hermite polynomials up to order $N$, the dimension of our finite basis is $M = \frac{1}{2}(N+1)(N+2)$. We define a reference vector for the $\mathscr{H}$ basis as 

\begin{equation}\label{eq:reference_vector}
	\mathbf{u} = \left( e_{0,0}, \cdots, e_{N,0}, \cdots, e_{0,N} \right).
\end{equation}

\noindent This is a vector-valued function of the phase space coordinates. For any vector $\mathbf{v}\in\mathbb{C}^M$, the contraction $ \mathbf{u}^\dagger\cdot\mathbf{v} $ returns a linear combination of the basis functions. 


When we choose the dimension $M$ for the matrix representation, we introduce coarse-graining by imposing a minimum length scale. The resultant scale is set by the maximum polynomial order $N$, and the model parameters that appear in the coordinate scaling (equation \ref{eq:coordinate_transformation}). The minimum scale can be quantified by the distance between adjacent roots in the highest order polynomial appearing in the basis functions, $e_{j,k}$. Note that the roots are not necessarily evenly spaced, so the effective resolution must be understood as a function of the phase space coordinates.

\subsection{Matrix Elements of $\hat{A}_0$}\label{section:A_0_elements}

In order to compute the spectrum of $\hat{A}_0$, we construct a  $M\times M$ matrix, which we denote $\bm{\mathsf{A}}_\mathsf{0}$. Its elements are obtained by computing the inner products, 

\begin{equation}\label{key}
	\left(\bm{\mathsf{A}}_\mathsf{0}\right)_{j',k'}^{j,k} = \langle E_{j',k'}, \hat{A}_0 E_{j,k}\rangle_*. 
\end{equation} 

\noindent Substituting $G\rightarrow E_{j,k}$, and $H\rightarrow H_0$ into equation \ref{eq:generator_definition}, and applying the definition of the $\mathscr{H}^*$ inner product in equation \ref{eq:dual_inner_product}, we obtain 

\begin{equation}\label{eq:shm_matrix_elements}
\begin{aligned}
	\left(\bm{\mathsf{A}}_\mathsf{0}\right)_{j',k'}^{j,k} = &\langle e_{j',k'}(q,p), [h_0(q,p),e_{j,k}(q,p)]\rangle.
\end{aligned}
\end{equation}

\noindent Expanding the Poisson bracket, and applying equation \ref{eq:hermite_property_1} we are left with the set of integrals,

\begin{equation}\label{key}
	\begin{aligned}
		\left(\bm{\mathsf{A}}_\mathsf{0}\right)_{j',k'}^{j,k} = &\kappa N_{j',k'}N_{j,k}\int_\mathcal{D} dqdp 
		f_0^2(q,p) \\
		& \times\bigg(
		 k q P_{j'}(q)P_j(q) P_{k'}(p) P_{k-1}(p) \\
		&- j p P_{k'}(p)P_k(p) P_{j'}(q)P_{j-1}(q)
		\bigg).
	\end{aligned}
\end{equation} 

\noindent The integrals are easier to evaluate if we first apply equation \ref{eq:hermite_property_2}. Doing so for $qP_j(q)$ in the first term and $pP_k(p)$ in the second gives

\begin{equation}\label{key}
	\begin{aligned}
		\left(\bm{\mathsf{A}}_\mathsf{0}\right)_{j',k'}^{j,k} = &\frac{\kappa}{2} N_{j',k'}N_{j,k}\int_\mathcal{D} dqdp 
		f_0^2(q,p) \\
		& \times\bigg(
		k  P_{j'}(q)P_{j+1}(q) P_{k'}(p) P_{k-1}(p) \\
		&- j  P_{k'}(p)P_{k+1}(p) P_{j'}(q)P_{j-1}(q)
		\bigg).
	\end{aligned}
\end{equation} 

\noindent From the orthogonality of the Hermite polynomials with respect to $f_0^2$, this reduces to 

\begin{equation}\label{key}
	\left(\bm{\mathsf{A}}_\mathsf{0}\right)_{j',k'}^{j,k} = \kappa \bigg(\sqrt{k(j+1)}\delta_{j'}^{j+1}\delta_{k'}^{k-1} - \sqrt{j(k+1)} \delta_{k'}^{k+1}\delta_{j'}^{j-1}\bigg).
\end{equation}

\noindent We organize the resulting $M^2$ real numbers into an $M\times M$ matrix according to the ordering of the $ e_{j,k} $ indices in the reference vector $\mathbf{u}$ (equation \ref{eq:reference_vector}). 

\subsection{Spectrum of $\hat{A}_0$}\label{section:A0_spectrum}

With a matrix representation of $\hat{A}_0$, we can compute a subset of its spectrum. Let  $\bm{\mathsf{A}}_\mathsf{0}\boldsymbol{\Psi}^{(0)} = \boldsymbol{\Psi}^{(0)}\boldsymbol{\Lambda}^{(0)}$ denote the eigendecomposition of $\bm{\mathsf{A}}_\mathsf{0}$, where 

\begin{equation}
	\boldsymbol{\Psi}^{(0)} = \begin{pmatrix}
		| & | &  &\\
		\boldsymbol{\psi}_1^{(0)} & \boldsymbol{\psi}_2^{(0)} & \cdots & \\
		| & | & &\
	\end{pmatrix}, 
	\ \ 
	\boldsymbol{\Lambda}^{(0)} = \begin{pmatrix}
		\lambda_1^{(0)} & 0 & \cdots \\
		0 & \lambda_2^{(0)} & \\
		\vdots & & \ddots\\
	\end{pmatrix}.
\end{equation}

\noindent Here, $\boldsymbol{\psi}_j^{(0)} \in \mathbb{C}^M$ and $\lambda_j^{(0)}\in\mathbb{C}$. We can transform the eigenvectors of $\bm{\mathsf{A}}_\mathsf{0}$ into eigenfunctions of $\hat{A}_0$ by taking their contraction with the reference vector $\mathbf{u}$. Denoting the eigenfunctions of $\hat{A}_0$ by $\psi_j^{(0)}(q,p)$, we have 

\begin{equation}\label{eq:eigenfunction}
	\psi_j^{(0)}(q,p) = \mathbf{u}^\dagger\cdot\boldsymbol{\psi}_j^{(0)}.
\end{equation}

In general, there can be a combination of real and imaginary eigenvalues. Complex eigenvalues come in conjugate pairs, as do their associated eigenvectors. All $\bm{\mathsf{A}}_\mathsf{0}$ eigenvalues $\lambda_j^{(0)}$ are determined by the characteristic polynomial,
	
\begin{equation}\label{eq:characteristic_polynomial}
	\mathrm{det}\left(\lambda^{(0)} \bm{\mathsf{I}} - \bm{\mathsf{A}}_\mathsf{0}\right) = 0,
\end{equation}	

where $\bm{\mathsf{I}}$ denotes the $M\times M$ identity matrix. We compute the  $\lambda_j^{(0)}$ by finding the roots of equation \ref{eq:characteristic_polynomial}. It is possible that some of these roots are repeated, and it is such repeated roots that we refer to as degenerate eigenvalues of $\bm{\mathsf{A}}_\mathsf{0}$. We will refer to the number of times a given eigenvalue is repeated as its multiplicity, denoted $d_j$.

The eigenvectors $\boldsymbol{\psi}_j^{(0)}$ associated with $\lambda_j^{(0)}$ are computed by solving,

\begin{equation}\label{eq:eigenvector_equation}
	 \left(\lambda_j^{(0)} \bm{\mathsf{I}}  - \bm{\mathsf{A}}_\mathsf{0}\right)\boldsymbol{\psi}_j^{(0)}= \mathbf{0}, 
\end{equation}

\noindent for $\boldsymbol{\psi}_j^{(0)}\neq\mathbf{0}$. We assume that all of the eigenvectors are linearly independent, including those associated with the degenerate eigenvalues. This means that associated with each eigenvalue $\lambda_{j}^{(0)}$, is a subspace of $\bm{\mathsf{A}}_\mathsf{0}$, spanned by the corresponding eigenvectors. For the non-degenerate eigenvalues, these subspaces are 1-dimensional, where as for the degenerate eigenvalues they have dimension equal to the degenerate eigenvalue multiplicity, $d_j$.

\begin{figure}
	\centering
	\includegraphics[width=\columnwidth]{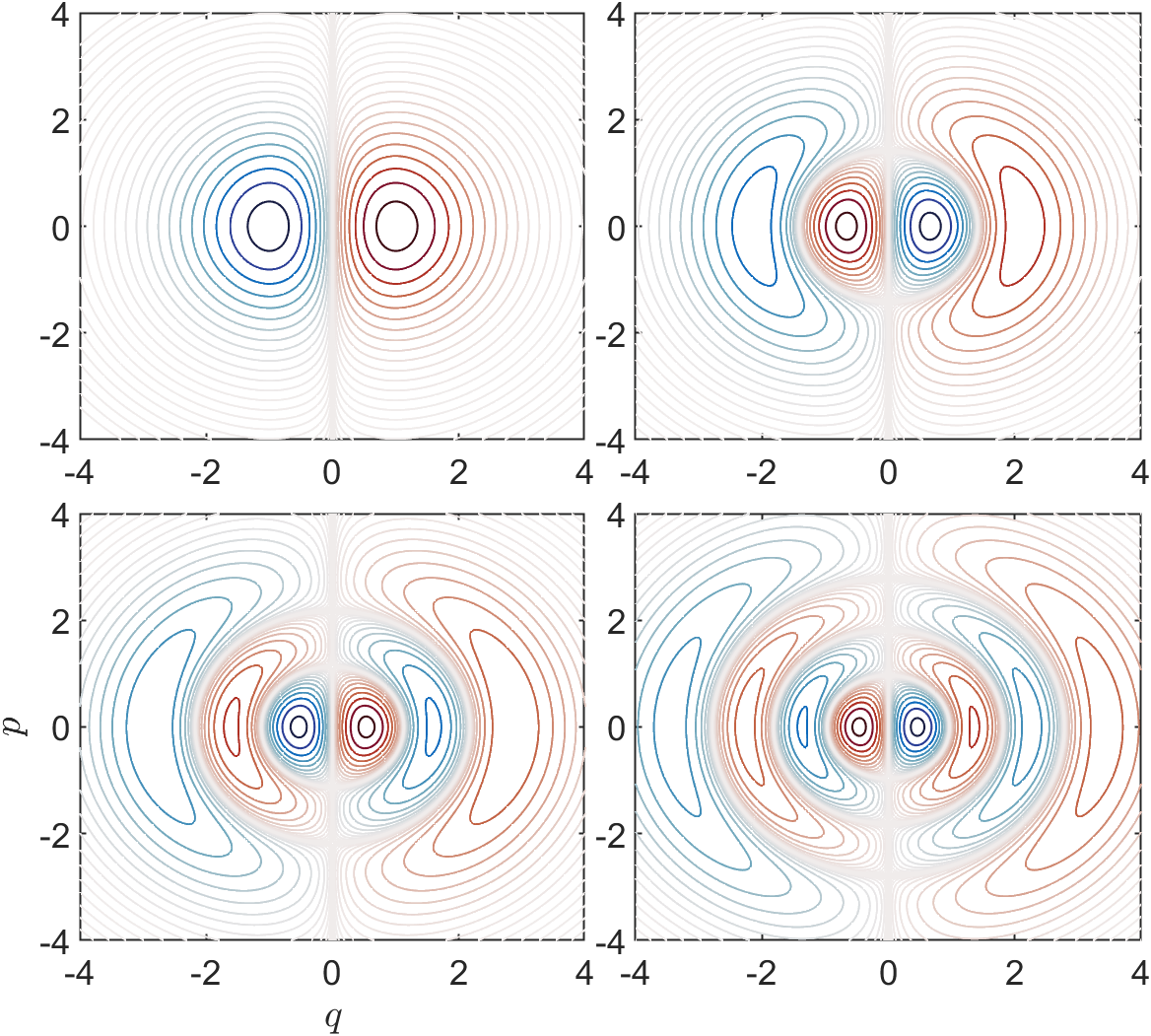}
	\caption{Level set contours of $\hat{A}_0$ eigenfunctions, denoted $\psi_{j}^{(0)}(q,p)$, for the degenerate subspace associated with $\lambda_j^{(0)}=-\mathrm{i}\kappa$. Each of these are linearly independent functions associated with their shared eigenvalue. We show only the real part of each, noting that the imaginary part in this case differs only by a rotation of $\pi/2$. This four dimensional subspace occurs for $N=8$.}
	\label{fig:shm_eigenfunctions_bending_mode}
\end{figure}

\begin{figure}
	\centering
	\includegraphics[width=\columnwidth]{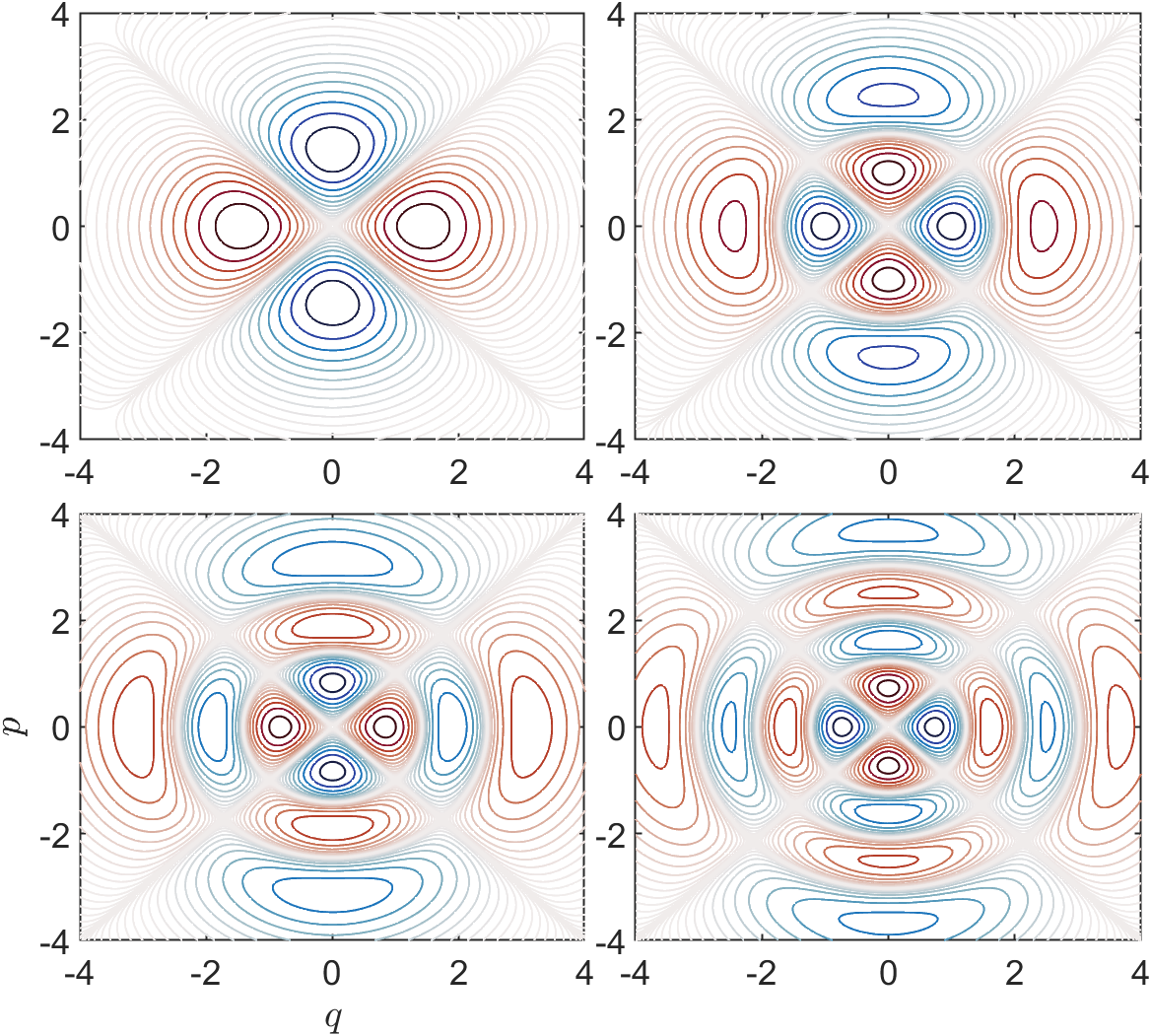}
	\caption{Same as Fig. \ref{fig:shm_eigenfunctions_bending_mode} but for $\lambda_j^{(0)}=-\mathrm{i}2\kappa$.}
	\label{fig:shm_eigenfunctions_breathing_mode}
\end{figure}

We show in Fig. \ref{fig:shm_eigenfunctions_bending_mode} and \ref{fig:shm_eigenfunctions_breathing_mode} the sets of linearly independent eigenfunctions associated with $\lambda_j^{(0)} = -i\kappa$ and $\lambda_j^{(0)} = -i2\kappa$ respectively. The former corresponds to the standard bending mode structure, and the latter the breathing mode. In each case, the presented function is the real part of one member of a conjugate pair. When the full complex functions are taken together with their conjugates, multiplication with the relevant complex exponentials produces real valued clockwise rotating bending and breathing patterns. 

\subsection{Dynamics of $f$ from $\hat{A}_0$}\label{section:moments_A0}

The dynamics of an arbitrary initial condition according to $h_0$ are determined by the spectrum of $\hat{A}_0$. Explicitly, $f$ is mapped to a new function $\hat{S}_0^tf$ at time $t$ by the flow of equation \ref{eq:CBE}, for $h = h_0$. We have from equation \ref{eq:f_expansion_t}, 

\begin{equation}\label{eq:sho_solution}
	\hat{S}_0^tf(q,p) = \sum_{j=1}^M\langle\psi_j^{(0)},f \rangle  \mathrm{e}^{\lambda_j^{\dagger(0)} t} \psi_j^{(0)}(q,p).
\end{equation} 

For the harmonic potential, the dynamics of $f$ are a rigid rotation of the initial condition about the origin. It is not immediately apparent from equation \ref{eq:sho_solution}, but we can verify that no deformation to the distribution occurs with the conserved quantities in the spectrum of $\hat{A}_0$. There are several eigenvectors of $\bm{\mathsf{A}}_\mathsf{0}$ associated with a zero eigenvalue. Each of these corresponds to a conserved quantity. By solving equation \ref{eq:eigenvector_equation} for $\lambda_j^{(0)}=0$, we find the sequence:

\begin{equation}\label{eq:constant_of_motion_sequence}
	\begin{aligned}
		\psi_1^{(0)}(q,p) &= e_{0,0},\\
		\psi_2^{(0)}(q,p) &= \tfrac{\sqrt{2}}{2}\left(e_{2,0} + e_{0,2}\right),\\
		\psi_3^{(0)}(q,p) &= \tfrac{1}{4}\left(\sqrt{6}e_{4,0} + 2e_{2,2} +\sqrt{6} e_{0,4}\right),\\
		\psi_4^{(0)}(q,p) &= \tfrac{1}{4}\left(\sqrt{5}e_{6,0} + \sqrt{3}e_{4,2} + \sqrt{3}e_{2,4} + \sqrt{5}e_{0,6}\right) ...\\
	\end{aligned}
\end{equation}

\noindent The first two correspond to conservation of phase space density and energy, and are themselves constants of motion in the sense that $[\psi_1^{(0)},h_0] = [\psi_2^{(0)},h_0] = 0$. The other eigenfunctions are not conserved in this sense. It is rather the corresponding functionals which are conserved. This can be understood in terms of the weighted statistical moments of $f$, defined as $F_{m,n}[f] = \langle f, f_0\  q^m p^n\rangle$. All of the zero-eigenvalue functions in equation \ref{eq:constant_of_motion_sequence} contain even polynomials of $q$ and $p$. This means that the corresponding functionals $\langle f, \psi_j^{(0)}\rangle$ are linear combinations of bivariate moments $F_{m,n}$ corresponding only to even powers in both variables. Such moments are variance, covariance, kurtosis, cokurtosis and so on. These moments quantify the character of symmetric properties of the distribution, ignoring asymmetries like mean and skew. This means that when the integrals over phase space are carried out, the result is the same regardless of where along the harmonic oscillator orbit the distribution is. Any deformation in $f$ will change the values of the integrals in this sequence. Their presence as zero-eigenvalue functionals in the spectrum of $\hat{A}_0$ restricts the dynamics to rigid rotation. 

\section{Anharmonic Potential}\label{section:anharmonic_potential}

With the harmonic oscillator spectrum calculated, we can now proceed to estimating the spectrum of $\hat{A}$. We begin by computing the matrix elements of $\hat{A}_4$.

\subsection{Matrix elements of $\hat{A}_4$}\label{section:A_4_elements}

The equivalent expression to equation \ref{eq:shm_matrix_elements} for the $\hat{A}_4$ matrix elements is obtained by making the substitutions $G\rightarrow E_{j,k}$, and $H\rightarrow  H_4$ in equation \ref{eq:generator_definition}. We have, 

\begin{equation}\label{eq:A_4_calculation_1}
\begin{aligned}
	\left(\bm{\mathsf{A}}_\mathsf{4}\right)_{j',k'}^{j,k} = & \Bigl\langle e_{j',k'}(q,p), \left[  \tfrac{4}{\kappa^4\beta^2}q^4,e_{j,k}(q,p)\right]\Bigr\rangle. \\
\end{aligned}
\end{equation}

\noindent We evaluate the Poisson bracket, applying equation \ref{eq:hermite_property_1} to the Hermite polynomial derivatives, and equation \ref{eq:hermite_property_2} to the $f_0$ derivatives. We are left with 

\begin{equation}\label{eq:A_4_calculation_2}
	\begin{aligned}
		\left(\bm{\mathsf{A}}_\mathsf{4}\right)_{j',k'}^{j,k} =& \frac{8}{\kappa^3\beta}N_{j',k'}N_{j,k}\int_\mathcal{D} dqdp 
		f_0^2(q,p)  q^3 P_{j'}(q)P_j(q)  \\
		&\times P_{k'}(p)\left(k P_{k-1}(p)  - \tfrac{1}{2}P_{k+1}(p)\right).
	\end{aligned}
\end{equation} 

\noindent To evaluate the these integrals for all indices, it is easiest to re-write $q^3P_j(q)$ strictly in terms of Hermite polynomials. Recursive application of equation \ref{eq:hermite_property_2} yields (for arbitrary $n\in\mathbb{N}$) an expansion of the form 

\begin{equation}\label{eq:hermite_property_3}
	x^nP_j(x) = \sum_{m=0}^{n}b_m(j)P_{j+n-2m}(x),
\end{equation}  

\noindent where the coefficients $b_m(j)$ are polynomials in $j$ of order $m$. Table \ref{tab:hermite_expansion_coef} contains a list of the relevant coefficients for both a quartic and sextic potential. To compute the contribution from $q^3P_j(q)$, let us define the integral 

\begin{equation}\label{eq:A_4_calculation_3}
	\begin{aligned}
		I_{j'}^j &= \int_{-\infty}^{+\infty} dq \ \mathrm{e}^{-q^2}q^3P_{j'}(q)P_j(q).
	\end{aligned}
\end{equation}

\noindent With this definition, the matrix elements of $\hat{A}_4$ (equation \ref{eq:A_4_calculation_2}) are expressed as

\begin{equation}\label{eq:A_4_calculation_5}
		\begin{aligned}
		\left(\bm{\mathsf{A}}_\mathsf{4}\right)_{j',k'}^{j,k} =& \frac{8}{Z^2\kappa^3\beta}N_{j',k'}N_{j,k}I_{j'}^j\int_{-\infty}^{+\infty} dp \ \mathrm{e}^{-p^2}\\
		  &\times P_{k'}(p)\left(k P_{k-1}(p)  - \tfrac{1}{2}P_{k+1}(p)\right),
	\end{aligned}
\end{equation}

\noindent where we have used $f_0^2 = Z^{-2}\mathrm{e}^{-q^2}\mathrm{e}^{-p^2}$.

Application of equation \ref{eq:hermite_property_3}, with the coefficients from the $n=3$ column of Table \ref{tab:hermite_expansion_coef} yields  

\begin{equation}\label{eq:A_4_calculation_4}
	\begin{aligned}
		I_{j'}^j 
		= \sum_{m=0}^{3}b_m(j) \int_{-\infty}^{+\infty} dq \ \mathrm{e}^{-q^2} P_{j'}(q)P_{j+3-2m}(q).
	\end{aligned}
\end{equation}

\noindent This is evaluated easily by the orthogonality of the Hermite polynomials, leaving 

\begin{equation}\label{key}
	I_{j'}^j = \sum_{m=0}^{3}b_m(j) \left(\sqrt{\pi} 2^{j+3-2m}(j+3-2m)!\right)^{\tfrac{1}{2}}\delta_{j'}^{j+3-2m}.
\end{equation}

\noindent We then substitute the evaluated $I_{j'}^j$ into equation \ref{eq:A_4_calculation_5}, and compute the remaining momentum space integral. We are left with 

\begin{equation}
	\begin{aligned}
\left(\bm{\mathsf{A}}_\mathsf{4}\right)_{j',k'}^{j,k} = &\frac{4\varphi}{\kappa^3\beta} \sum_{m=0}^{3}b_m(j) N_{j,k} \delta_{j'}^{j+3-2m} \\
&\times \left(\frac{2k\delta_{k'}^{k-1}}{N_{j+3-2m,k-1}} - \frac{\delta_{k'}^{k+1}}{N_{j+3-2m,k+1}}\right).
	\end{aligned}
\end{equation}

\noindent Again, we organize the resulting values into an $M\times M$ matrix according to the ordering of indices in $\mathbf{u}$.

\begin{table}
	\centering
	\begin{tabular}{|c| c| c|}
		\hline 
		$m$ & $b_m$ $(n=3)$ & $b_m$ $(n=5)$ \\
		\hline
		0 & $2^{-3}$& $2^{-5}$\\
		1 & $\frac{3}{4}(j+1)$& $\frac{1}{2}(j+\frac{5}{4})$ \\
		2 & $\frac{3}{2}j^2$& $\frac{5}{2}(\frac{1}{2}j^2 + j + \frac{3}{4})$\\
		3 & $j!/(j-3)!$& $\frac{5}{2}j(j^2 + \frac{1}{2})$\\
		4 & & $5j(\frac{1}{2}j^3-2j^2+\frac{5}{2}j-1)$\\
		5 & & $j!/(j-5)!$\\
		\hline
	\end{tabular} 
	\caption{Coefficients for Hermite polynomial expansion of matrix element integrand as in equation \ref{eq:hermite_property_3}.}
	\label{tab:hermite_expansion_coef}
\end{table}

\subsection{Spectrum of $\hat{A}$}\label{section:spectrum_of_A_4}

To apply the ordinary differential equation solution for $\hat{U}^t$, and compute $\hat{S}^tf$ from equation \ref{eq:f_expansion_t}, we need the spectrum of $\hat{A}$. The matrix

\begin{equation}
\bm{\mathsf{A}} =	\bm{\mathsf{A}}_\mathsf{0} + \varphi\bm{\mathsf{A}}_\mathsf{4},
\end{equation}

\noindent does not admit a diagonalization directly, so we adopt a perturbative treatment. We follow a standard procedure for analysis of the Schr{\"o}dinger equation in QM (for an introductory description, see for example \cite{griffiths2018}). Suppose that the eigenvalues and eigenvectors of $\bm{\mathsf{A}}$ can be written as a power series in the parameter $\varphi$. That is,

\begin{equation}\label{eq:perturbation_power_series_single_index}
	\begin{aligned}
		\lambda_{j}(\varphi) &= \lambda_{j}^{(0)} + \varphi\lambda_{j}^{(1)} + \mathcal{O}(\varphi^2), \\
		\boldsymbol{\psi}_{j}(\varphi) &= \boldsymbol{\psi}_{j}^{(0)} + \varphi\boldsymbol{\psi}_{j}^{(1)} + \mathcal{O}(\varphi^2).
	\end{aligned}
\end{equation}

\noindent We assume that these quantities satisfy the eigenvalue problem, 

\begin{equation}\label{eq:perturbed_eigenvalue_problem}
	\left(\bm{\mathsf{A}}_\mathsf{0}+\varphi\bm{\mathsf{A}}_\mathsf{4}\right)\boldsymbol{\psi}_j(\varphi) = \lambda_j(\varphi)\boldsymbol{\psi}_j(\varphi).
\end{equation}

\noindent Substituting the expressions in equation \ref{eq:perturbation_power_series_single_index} into \ref{eq:perturbed_eigenvalue_problem} and equating terms of equal power in $\varphi$, we obtain a sequence of equations relating the power series coefficients and the operators $\bm{\mathsf{A}}_\mathsf{0}$ and $\bm{\mathsf{A}}_\mathsf{4}$. The zero order equation is the eigenvalue problem for $\bm{\mathsf{A}}_\mathsf{0}$, stated in Section \ref{section:A0_spectrum}. The first order equation is 

\begin{equation}\label{eq:degenerate_PT_equal_powers_single_index}
	\begin{aligned}
		\left(\bm{\mathsf{A}}_\mathsf{0} - \lambda_{j}^{(0)}\right)\boldsymbol{\psi}_{j}^{(1)} &= \left(\lambda^{(1)}_{j} - \bm{\mathsf{A}}_\mathsf{4}\right)\boldsymbol{\psi}_{j}^{(0)}.
	\end{aligned}
\end{equation}

We assume that the basis formed by the eigenvectors of $\bm{\mathsf{A}}_\mathsf{0}$ spans the eigenspace of $\bm{\mathsf{A}}$. That is,

\begin{equation} \label{eq:eigenvector_expansion}
	\boldsymbol{\psi}_j(\varphi) = \sum_{k=1}^{M} \left({\boldsymbol{\psi}_k^{(0)}}^\dagger\cdot\boldsymbol{\psi}_j(\varphi)\right)\boldsymbol{\psi}_k^{(0)}.
\end{equation}

\noindent Note that $ {\boldsymbol{\psi}_k^{(0)}}^\dagger\cdot\boldsymbol{\psi}_j(\varphi) $ indicates a dot product in $\mathbb{C}^M$ and the result is scalar. Substituting  $\boldsymbol{\psi}_j(\varphi)$ as in equation \ref{eq:perturbation_power_series_single_index} into equation \ref{eq:eigenvector_expansion}, we express the first order correction as

\begin{equation}\label{eq:non_degenerate_eigenvector_correction_1}
	\boldsymbol{\psi}_j^{(1)} = \sum_{k=1}^{M} \left({\boldsymbol{\psi}_k^{(0)}}^\dagger\cdot\boldsymbol{\psi}_j^{(1)}\right)\boldsymbol{\psi}_k^{(0)}.
\end{equation} 

\noindent In order to compute this correction we need an expression for the coefficients, ${\boldsymbol{\psi}_k^{(0)}}^\dagger\cdot\boldsymbol{\psi}_j^{(1)} $. We proceed by contracting equation \ref{eq:degenerate_PT_equal_powers_single_index} with ${\boldsymbol{\psi}_k^{(0)}}^\dagger$. This yields 

\begin{equation}
	\left(\lambda_k^{(0)}-\lambda_j^{(0)}\right){\boldsymbol{\psi}_k^{(0)}}^\dagger\cdot\boldsymbol{\psi}^{(1)}_j = \lambda_j^{(1)}\delta_j^k - {\boldsymbol{\psi}_k^{(0)}}^\dagger\cdot\bm{\mathsf{A}}_\mathsf{4}\boldsymbol{\psi}_j^{(0)}.
\end{equation}

\noindent For $k=j$, the left hand side is zero, and we have the first order correction to the eigenvalues, 

\begin{equation}
	\lambda_j^{(1)} = {\boldsymbol{\psi}_j^{(0)}}^\dagger\cdot\bm{\mathsf{A}}_\mathsf{4}\boldsymbol{\psi}_j^{(0)}.
\end{equation}

\noindent For $k\neq j$, $\delta_j^k=0$ and we obtain the coefficients we need for equation \ref{eq:non_degenerate_eigenvector_correction_1}. That is,

\begin{equation}\label{eq:non_degenerate_eigenvector_correction_2}
	{\boldsymbol{\psi}_k^{(0)}}^\dagger\cdot\boldsymbol{\psi}^{(1)}_j = \frac{{\boldsymbol{\psi}_k^{(0)}}^\dagger\cdot\bm{\mathsf{A}}_\mathsf{4}\boldsymbol{\psi}_j^{(0)}}{\lambda_j^{(0)}-\lambda_k^{(0)}}.
\end{equation}

\noindent Summing over these coefficients as in equation \ref{eq:non_degenerate_eigenvector_correction_1} yields the $\mathcal{O}(\varphi)$  correction to the eigenvectors of $\bm{\mathsf{A}}$. Explicitly, the first order correction to the non-degenerate eigenvectors of $\bm{\mathsf{A}}$ is 

\begin{equation}\label{eq:non_degenerate_eigenvector_correction}
	\boldsymbol{\psi}_j^{(1)} = \sum_{k=1}^M 
	\frac{{\boldsymbol{\psi}_k^{(0)}}^\dagger\cdot\bm{\mathsf{A}}_\mathsf{4}\boldsymbol{\psi}_j^{(0)}}{\lambda_j^{(0)}-\lambda_k^{(0)}}
	\boldsymbol{\psi}_k^{(0)}.
\end{equation}

We can use the corrections computed here as-is for the non-degenerate subset of the $\bm{\mathsf{A}}_\mathsf{0}$ spectrum, where $\lambda_j^{(0)}\neq \lambda_k^{(0)}$ if $j\neq k$. The situation is more complicated if $\lambda_j^{(0)}= \lambda_k^{(0)}$. As discussed in Section \ref{section:A0_spectrum}, the spectrum of $\bm{\mathsf{A}}_\mathsf{0}$ in general contains degenerate subspaces, where there are multiple distinct eigenvectors associated with the same eigenvalue (see for example Fig. \ref{fig:shm_eigenfunctions_bending_mode} and \ref{fig:shm_eigenfunctions_breathing_mode}). This poses an issue for the correction in equation \ref{eq:non_degenerate_eigenvector_correction}, where we must sum over all of the linearly independent eigenvectors. If $j$ and $k$ are such that both eigenvectors lie within a degenerate subspace, the denominator here is zero. To proceed, we treat the $\bm{\mathsf{A}}_\mathsf{4}$ contribution separately for each degenerate subspace.


\subsection{Treatment of degenerate eigenvalues}\label{section:degenerate_subspace} 

We begin by adding a second index label to the eigenvalues of $\bm{\mathsf{A}}_\mathsf{0}$. That is, let $\lambda_{j,k}^{(0)}$ denote the $k$th repetition of the $j$th unique eigenvalue of $\bm{\mathsf{A}}_\mathsf{0}$. The same indexing scheme is applied to the eigenvectors. As stated in Section \ref{section:A0_spectrum}, all eigenvectors corresponding to repeated eigenvalues are themselves linearly independent. So although we have $\lambda_{j,k}^{(0)} = \lambda_{j,k'}^{(0)}$, for the eigenvectors,  $\boldsymbol{\psi}_{j,k}^{(0)}\neq\boldsymbol{\psi}_{j,k'}^{(0)}$. 

For any fixed $j$, the $\boldsymbol{\psi}_{j,k}^{(0)}$ are linearly independent and span a subspace of the $\bm{\mathsf{A}}_\mathsf{0}$ eigenspace corresponding to a single eigenvalue $\lambda_j^{(0)}$. Any linear combination of these vectors is itself an eigenvector of $\bm{\mathsf{A}}_\mathsf{0}$, with eigenvalue $\lambda_j^{(0)}$. We can make use of this property in determining the $\bm{\mathsf{A}}_\mathsf{4}$ contribution. We begin by looking at the spectrum of $\bm{\mathsf{A}}_\mathsf{4}$ when it is projected onto each degenerate subspace of $\bm{\mathsf{A}}_\mathsf{0}$ separately. This process is as follows. 

For each eigenvalue $\lambda_j^{(0)}$ with multiplicity $d_j>1$, we define the subspace basis as

\begin{equation}\label{eq:subspace_basis}
	\boldsymbol{\Psi}_j^{(0)} = \begin{pmatrix}
		| & & | \\
		\boldsymbol{\psi}_{j,1}^{(0)} & \cdots &  \boldsymbol{\psi}_{j,d_j}^{(0)}\\
		| & & |
	\end{pmatrix}.
\end{equation}

\noindent We then compute the matrix elements of $\hat{A}_4$ projected onto the degenerate subspace. Denoting the $j$th subspace projection $\tilde{\bm{\mathsf{A}}}_\mathsf{4}^j$, we have 

\begin{equation}\label{key}
	(\tilde{\bm{\mathsf{A}}}_\mathsf{4}^j)_k^{k'} = {\boldsymbol{\psi}_{j,k}^{(0)}}^\dagger \cdot \bm{\mathsf{A}}_\mathsf{4} \boldsymbol{\psi}_{j,k'}^{(0)}.
\end{equation}

\noindent Note that the projected matrices are $d_j\times d_j $ rather than $M\times M$. With the projected generator matrix, we compute its eigenvalues, $\lambda^{(1)}_{j,k}\in \mathbb{C}$ and eigenvectors, $\boldsymbol{\xi}_{j,k}\in\mathbb{C}^{d_j}$. These satisfy 

\begin{equation}\label{key}
	\tilde{\bm{\mathsf{A}}}_\mathsf{4}^j\boldsymbol{\xi}_{j,k} = \lambda^{(1)}_{j,k}\boldsymbol{\xi}_{j,k}.
\end{equation}

\subsection{Simultaneous eigenvectors of $\hat{A}_0$ and $\hat{A}_4$}\label{section:simultaneous_eigenvectors}

The $\boldsymbol{\xi}_{j,k}$ comprise weights for linear combinations of the degenerate subspace basis vectors $\boldsymbol{\psi}^{(0)}_{j,k}$. Contracting with the subspace basis $\boldsymbol{\Psi}_j^{(0)}$ in equation \ref{eq:subspace_basis}, we can get back vectors in $\mathbb{C}^M$. Explicitly, we write

\begin{equation}\label{eq:simultaneous_eigenvectors}
	\tilde{\boldsymbol{\psi}}_{j,k}^{(0)} = \boldsymbol{\Psi}^{(0)}_j\boldsymbol{\xi}_{j,k}, \ \ 1\leq k \leq d_j.
\end{equation}

\noindent Each of the $\tilde{\boldsymbol{\psi}}_{j,k}^{(0)}$ are eigenvectors of both the projected $\bm{\mathsf{A}}_\mathsf{4}$, with eigenvalue $\lambda^{(1)}_{j,k}$, and of $\bm{\mathsf{A}}_\mathsf{0}$ with eigenvalue $\lambda_j^{(0)}$. Because of this, we can replace the $d_j$ original eigenvectors of $\bm{\mathsf{A}}_\mathsf{0}$ making up the $j$th degenerate subspace with the new $\tilde{\boldsymbol{\psi}}_{j,k}^{(0)}$. 

We need these new eigenvectors to resolve an ambiguous definition of the unperturbed $\boldsymbol{\psi}_{j,k}^{(0)}$. If we did not do this, the transition between $\varphi=0$ and $\varphi>0$ could constitute a discontinuous change in the eigenvectors. We must avoid this to assure that we have a smooth variation of the eigenvectors with respect to change in the perturbation parameter $\varphi$. Otherwise, the original power series assumption (equation \ref{eq:perturbation_power_series_single_index}) would be invalid. From this point on, we replace all of the degenerate subspace eigenvectors, $\boldsymbol{\psi}_{j,k}^{(0)}$ with the new vectors we just found, $\tilde{\boldsymbol{\psi}}^{(0)}_{j,k}$. For notational simplicity, we drop the tilde. Any time from this point on that we refer to the degenerate subspace vectors, it should be assumed that they are the ones defined by equation \ref{eq:simultaneous_eigenvectors} that diagonalize $\bm{\mathsf{A}}_\mathsf{4}$ in their respective subspaces. The eigenfunctions are obtained from the eigenvectors in the same way as for the harmonic oscillator, using an analog to equation \ref{eq:eigenfunction}. 

\subsection{Degenerate Subspace Corrections}

To compute the $\mathcal{O}(\varphi)$ corrections for the degenerate eigenvalues and eigenvectors, we follow a similar procedure to Section \ref{section:A0_spectrum}, carrying out the necessary algebra in Appendix \ref{A:degenerate_perturbation}. To summarize, we again assume a power series in $\varphi$ for both the eigenvalues and eigenvectors of $\bm{\mathsf{A}}$ (equation \ref{eq:perturbation_power_series}), and then compute the coefficients of the $\mathcal{O}(\varphi)$ terms in these series.

\subsubsection{Eigenvalues of $\hat{A}$}\label{section:eigenvalues_of_A}
	
The expression for the first order corrected eigenvalues is the same as it was in the non-degenerate case, just with the double index on the degenerate subsets. Explicitly,

\begin{equation}\label{eq:degenerate_values_series}
	\lambda_{j,k}(\varphi) = \lambda_j^{(0)} + \varphi \lambda_{j,k}^{(1)} + \mathcal{O}(\varphi^2),
\end{equation}

\noindent where $\lambda_{j,k}^{(1)}$ is given by equation \ref{eq:degenerate_eigenvalues_correction_1}, derived in Appendix \ref{A:degenerate_eigenvalues}. This is obtained by diagonalizing $\bm{\mathsf{A}}_\mathsf{4}$ in the $j$th degenerate subspace as in Section \ref{section:degenerate_subspace}. Here, the eigenvalues of $\bm{\mathsf{A}}_\mathsf{4}$ are distinct within any given degenerate subspace $\boldsymbol{\Psi}_j^{(0)}$. That is, $ \lambda^{(1)}_{j,k}\neq\lambda^{(1)}_{j,k'} $ for $ k\neq k' $. This means that the degenerate eigenvalues of $\bm{\mathsf{A}}_\mathsf{0}$ have been split by $\bm{\mathsf{A}}_\mathsf{4}$. If this is not true, a higher order expansion in $\varphi $ is required.

\begin{figure}
	\centering
	\includegraphics[width=\columnwidth]{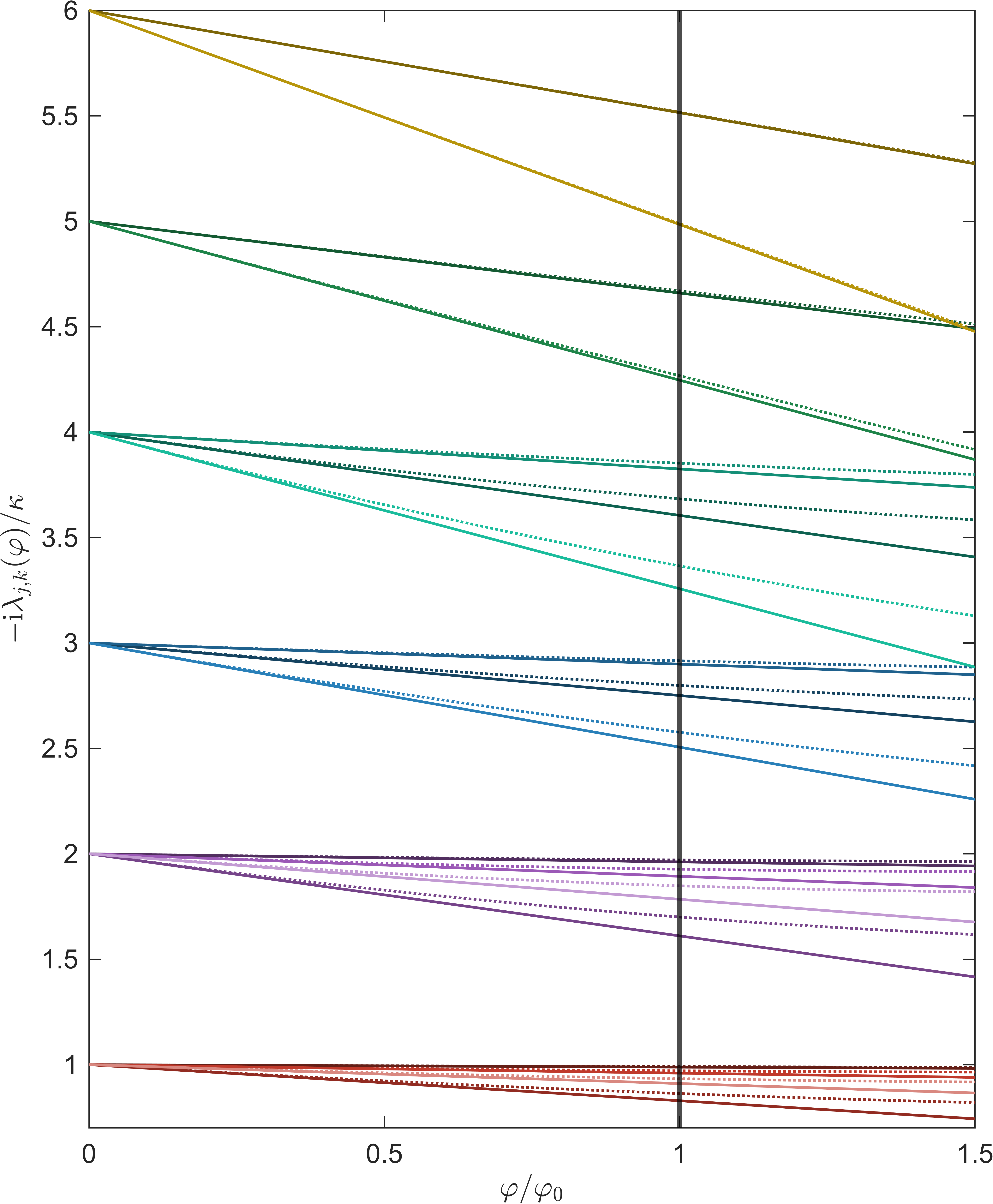}
	\caption{Corrected eigenvalues of $\hat{A}$ corresponding to the degenerate subspaces of $\hat{A}_0$, as a function of $\varphi$. The horizontal axis is scaled by $\varphi_0 = -\frac{\beta}{12}(\frac{\kappa}{2})^4$. The solid and dotted lines correspond to the $\mathcal{O}(\varphi)$ and $\mathcal{O}(\varphi^2)$ corrections respectively. Curves of matching hue (and diverging from the same point at $\varphi=0$) lie within the same degenerate subspace. The first order corrections are given by equation \ref{eq:degenerate_eigenvalues_correction_1}, and the second order corrections are computed according to equation \ref{eq:second_order_eigenvalues}. This particular set of eigenvalues is for $N=8$.}
	\label{fig:split_eigenvalues}
\end{figure}

We show in Fig. \ref{fig:split_eigenvalues} the corrected eigenvalues corresponding to half of the degenerate subspaces (the values shown are conjugate partners to those in the omitted half). The $-\mathrm{i}\lambda_{j,k}(\varphi)/\kappa$ curves serve to illustrate the splitting of the degenerate subset of the $\hat{A}_0$ spectrum. The degenerate subspace dimension decreases as $|\lambda_j^{(0)}|$ increases. The curves corresponding to a single degenerate eigenvalue of $\hat{A}_0 $ converge at $\varphi = 0$, but possess distinct values otherwise. The spread of the $\lambda_{j,k}(\varphi)$ for fixed $j$, increases with $\varphi$. This corresponds to the strength of the anharmonic contribution to the potential. It is this distance between eigenvalues of $\hat{A}$ within a given degenerate subspace that drives differential rotation in the distribution function. The vertical line at $\varphi/\varphi_0 = 1$ indicates the chosen value of $\varphi$ used in our example calculations.

\subsubsection{Eigenvectors of $\hat{A}$}\label{section:eigenvectors_of_A}

The eigenvectors of $\bm{\mathsf{A}}$ are expressed as the power series,

\begin{equation}
	\boldsymbol{\psi}_{j,k}(\varphi) = \boldsymbol{\psi}_{j,k}^{(0)} + \varphi\boldsymbol{\psi}_{j,k}^{(1)} + \mathcal{O}(\varphi^2).
\end{equation}

\noindent The first order correction $\boldsymbol{\psi}_{j,k}^{(1)}$ is derived in Appendix \ref{A:degenerate_eigenvectors}, and can ultimately be written as

\begin{equation}\label{eq:degenerate_eigenvectors_correction_1}
	\boldsymbol{\psi}_{j,k}^{(1)} = \sum_{m\neq j }^M \frac{{\boldsymbol{\psi}_{m}^{(0)}}^\dagger\cdot \bm{\mathsf{A}}_4\boldsymbol{\psi}_{j,k}^{(0)}}{\lambda_{m}^{(0)} - \lambda_{j}^{(0)}}\left(
	\sum_{l\neq k }^{d_j}
	\frac{{\boldsymbol{\psi}_{j,l}^{(0)}}^\dagger\cdot\bm{\mathsf{A}}_4 
		\boldsymbol{\psi}_m^{(0)}}
	{\lambda_{j,l}^{(1)} - \lambda_{j,k}^{(1)}}\boldsymbol{\psi}_{j,l}^{(0)}
	-	\boldsymbol{\psi}_m^{(0)} 
	\right).
\end{equation} 

\noindent As in Section \ref{section:A0_spectrum}, we can transform the eigenvectors of $\bm{\mathsf{A}}$ into eigenfunctions of $\hat{A}$ by taking their contraction with the reference vector, $\mathbf{u}$. We have 

\begin{equation}\label{eq:first_order_degenerate_eigenfunctions}
	\psi_{j,k}(q,p;\varphi) = \mathbf{u}^\dagger \cdot \boldsymbol{\psi}_{j,k}(\varphi).
\end{equation}

\begin{figure}
	\centering
	\includegraphics[width=\columnwidth]{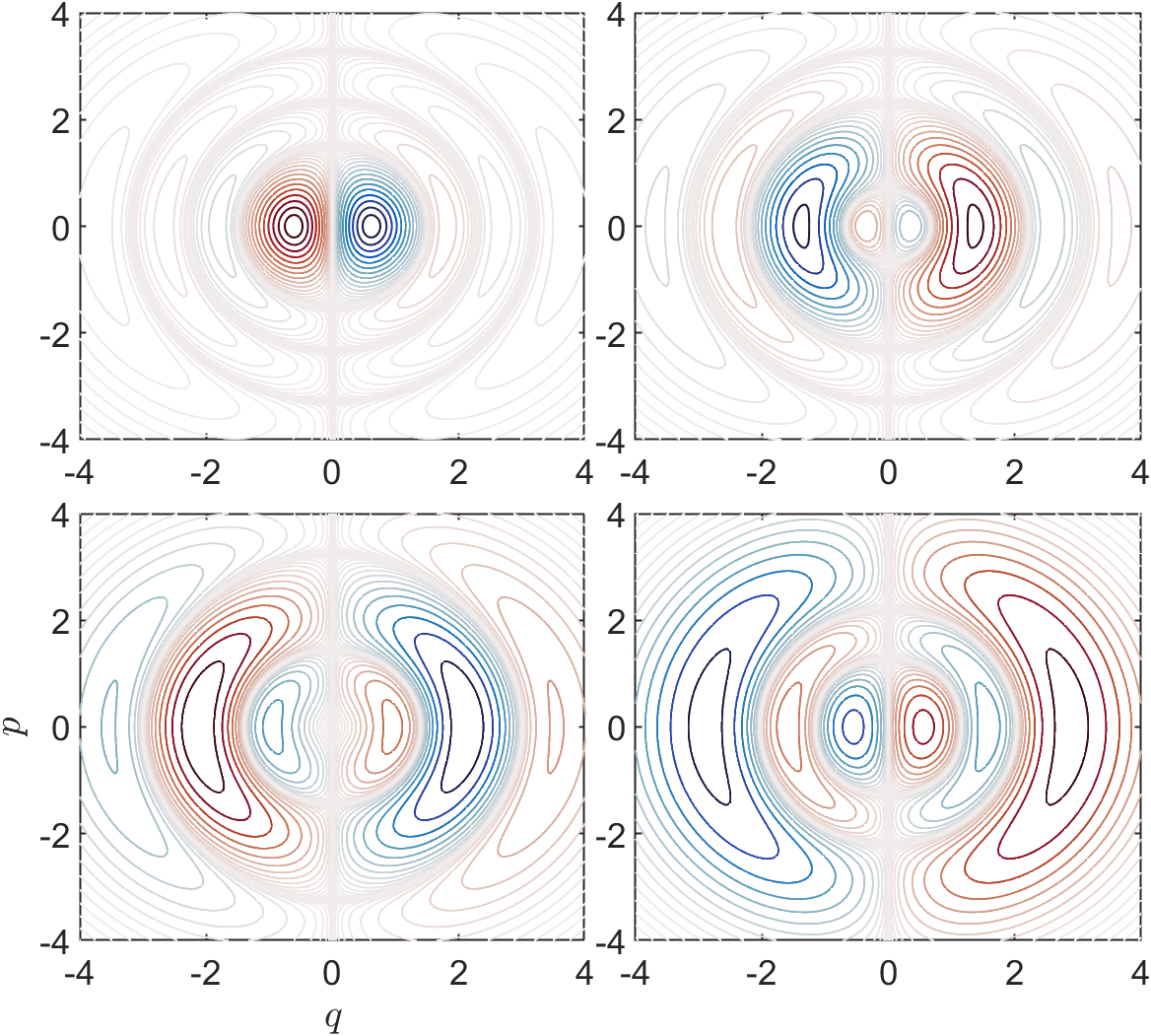}
	\caption{Approximate eigenfunctions of $\hat{A}$, denoted $\psi_{j,k}(q,p;\varphi)$, associated with the $\lambda^{(0)}_j=-\mathrm{i}\kappa$ degenerate subspace (bending mode). These are specified by equations  \ref{eq:degenerate_eigenvectors_correction_1} and \ref{eq:first_order_degenerate_eigenfunctions}. Going from left to right and top to bottom, the panels are sorted from fastest to slowest rate of rotation. We show only the real part of each, noting that the imaginary parts in this case differ only by a rotation. This particular set of eigenfunctions is for $N=8$.}
	\label{fig:degenerate_eigenfunctions_1}
\end{figure}

We show in Fig. \ref{fig:degenerate_eigenfunctions_1} an example set of $\hat{A}$ eigenfunctions associated with $\lambda_j^{(0)}=-\mathrm{i}\kappa$. As discussed in Section \ref{section:A0_spectrum}, this is the eigenvalue that controls the rotation of the simple bending mode in Fig. \ref{fig:shm_eigenfunctions_bending_mode}. The $\hat{A}$ eigenfunctions contain more sign changes along a radial path from the origin than those of the harmonic oscillator. This facilitates a segmenting of phase space density with distance from the origin (proportional to the action in this case). Each of these structures possesses a different orbital frequency, corresponding to the four diverging solid red lines starting at $-\mathrm{i}\lambda_{j,k}(\varphi)/\kappa=1$ in Fig. \ref{fig:split_eigenvalues}. The relative spacing of these lines determines the relative rates of rotation, and therefore the relative phase of the structures in Fig. \ref{fig:degenerate_eigenfunctions_1}, as time progresses. The combination of these two factors causes differential rotation in the distribution function, producing the spiral structure characteristic of phase mixing. The same general idea applies to the other degenerate subspace eigenfunctions, although there are more complicated structures present in the larger $|\lambda_j^{(0)}|$ eigenfunctions. 

To illustrate the importance of degenerate subspaces to the mixing process, we consider the sum over eigenvectors within a given subspace. Let us define the $j$th subspace sum as

\begin{equation}\label{eq:subspace_sum}
	\psi_j^{\Sigma}(q,p,t) = \sum_{k=1}^{d_j} \langle \psi_{j,k}, f\rangle\psi_{j,k}(q,p;\varphi)\mathrm{e}^{\lambda_{j,k}^\dagger(\varphi)t}.
\end{equation}

\begin{figure}
	\centering
	\includegraphics[width=\columnwidth]{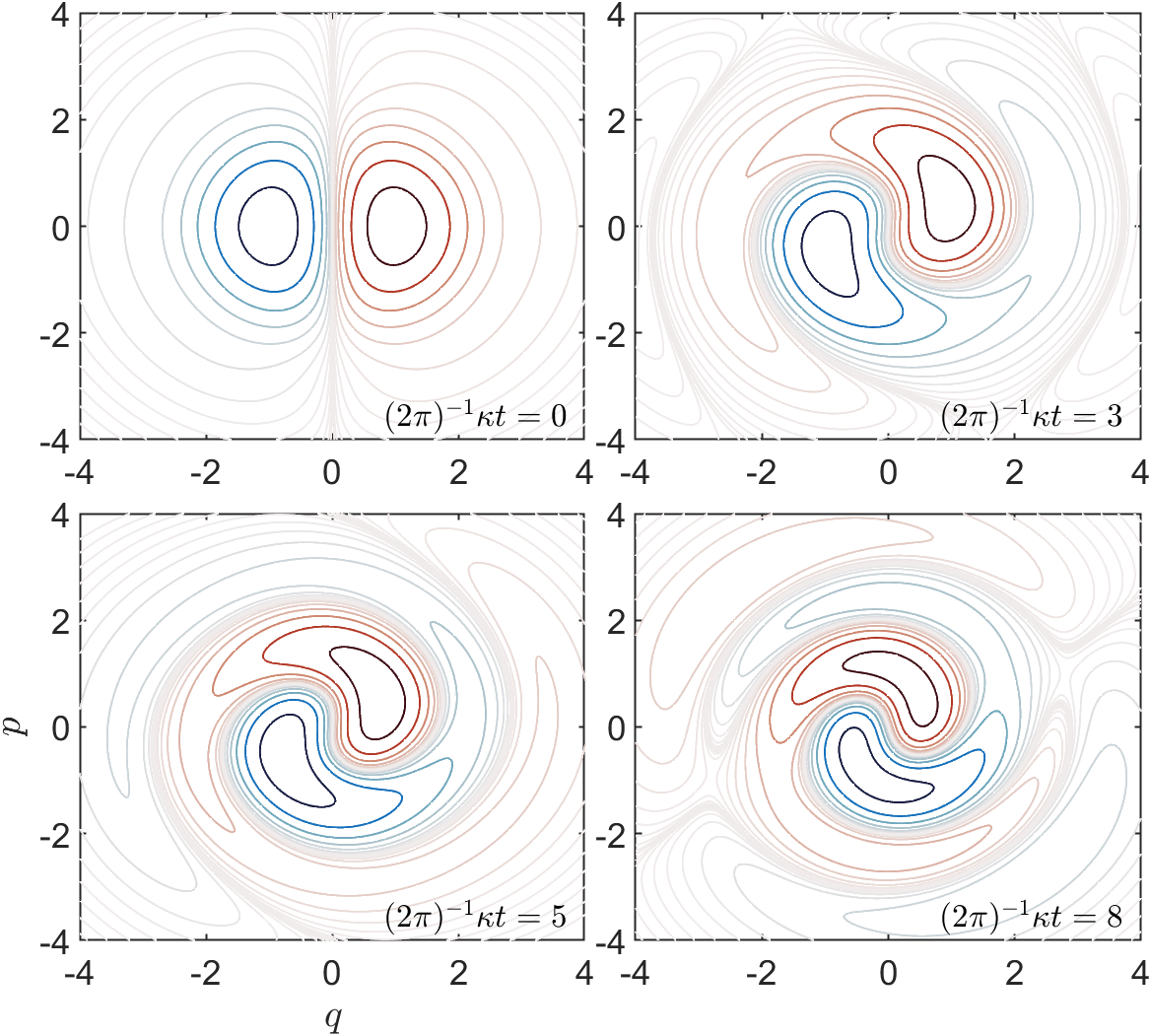}
	\caption{Time evolution of the $\lambda_j^{(0)}=-\mathrm{i}\kappa$ degenerate subspace with first order corrections from $\hat{A}_4$ to the eigenvalues and eigenvectors. This is the real part of the subspace sum $\psi_j^\Sigma(q,p,t)$ corresponding to the set of eigenfunctions shown in Fig. \ref{fig:degenerate_eigenfunctions_1}, calculated according to equation \ref{eq:subspace_sum}.}
	\label{fig:degenerate_eigenfunctions_subspace_contribution}
\end{figure}

\noindent We show this quantity in Fig. \ref{fig:degenerate_eigenfunctions_subspace_contribution}, with the particular $j$ value corresponding to the index of the $\lambda_j^{(0)}=-\mathrm{i}\kappa$ subspace. The initial condition is $f_0(q,p+\delta p)$, with $\delta p=0.3$. At $t=0$, the split eigenvalues have no effect, and the net contribution from the subspace is the bending mode. As time progresses, the eigenfunctions in Fig. \ref{fig:degenerate_eigenfunctions_1} rotate increasingly out of phase with each other, deforming the bending mode structure into interlocked positive and negative one-armed spirals.

We have observed so far that particular types of structures correspond to degenerate subspaces of the harmonic oscillator spectrum. To reiterate, these subspaces are sets of linearly independent eigenfunctions of $\hat{A}_0$, all associated with the same eigenvalue. When the anharmonic potential contribution is added, the degenerate eigenvalues split, but the correspondence between particular structures (bending, breathing, etc.) and the degenerate subspaces remains. The subspace dimension determines the complexity of representable structure, and taken in combination with the properties of the contained eigenfunctions (spacing of their roots), sets the minimum length scale of that structure.

\subsection{Dynamics of $f$ from $\hat{A}$}

With an estimated spectrum of the full generator, $\hat{A}$, we can compute the distribution function at time $t$. Separating the sum in equation \ref{eq:f_expansion_t} into the distinct contributions from our perturbative analysis we have

\begin{equation}\label{eq:DF_solution_anharmonic}
	\begin{aligned}
		\hat{S}^tf(q,p) =  \sum_{j=1}^{M-D} \langle\psi_j,f \rangle\psi_j(q,p;\varphi) \mathrm{e}^{\lambda_j^\dagger(\varphi) t} 
		+ \sum_{j>M-D}^M\psi_j^{\Sigma}(q,p,t).
	\end{aligned}
\end{equation}

\noindent On the right hand side from left to right these are, the non-degenerate subspace of the spectrum of $\hat{A}$, and the total contribution from the $D$ degenerate subspaces. 

\begin{figure*}
	\centering
	\includegraphics[width=2\columnwidth]{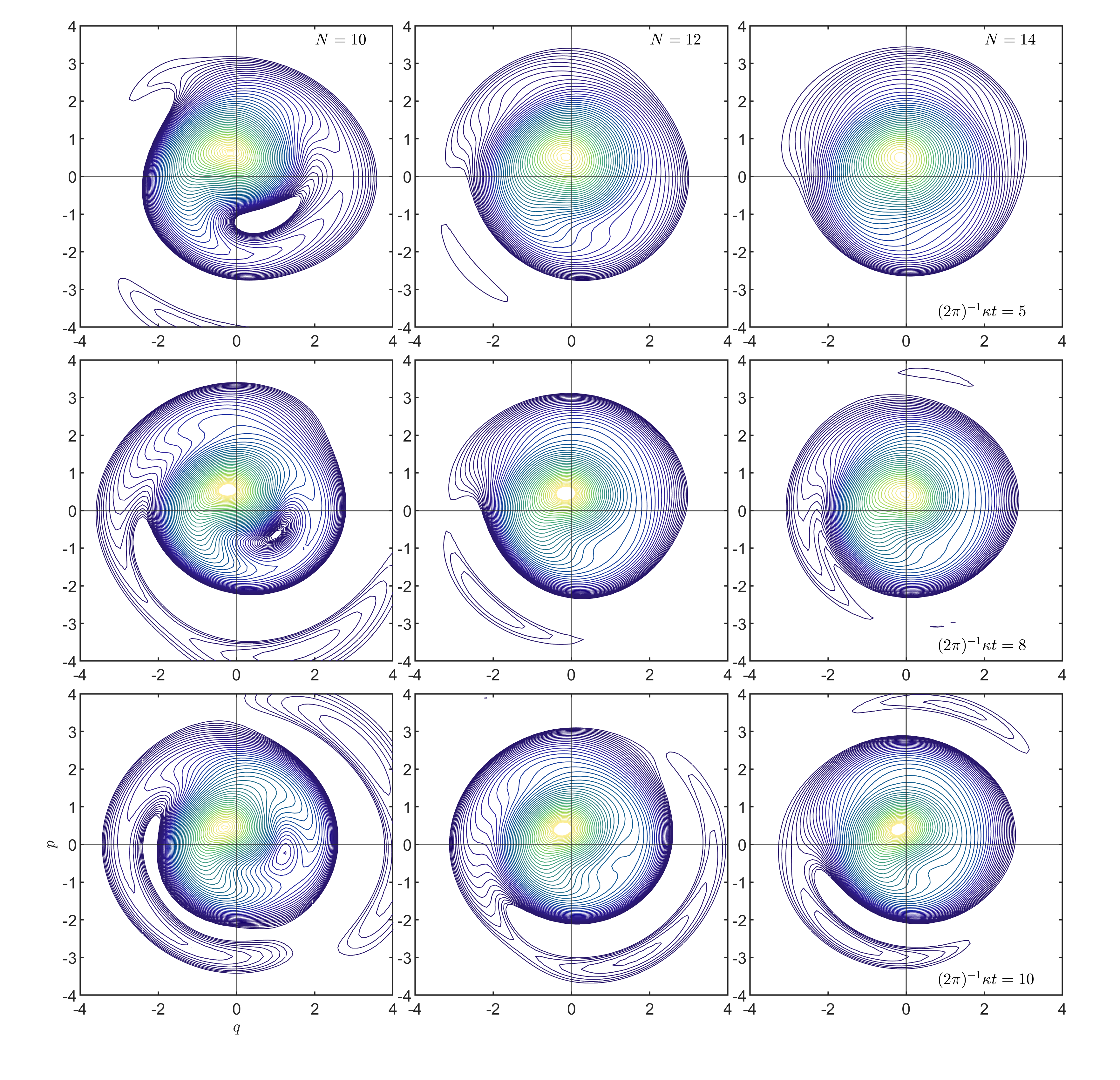}
	\caption{Snapshots of the distribution function $\hat{S}^tf$, computed from equation \ref{eq:DF_solution_anharmonic}. From left to right, $N=10,12,14$. From top to bottom, $(2\pi)^{-1}\kappa t = 5,8,10$. In all panels, the $j$th contour value is given by $\mathrm{max}\{f_0\}\mathrm{e}^{-\frac{\beta}{2}x_j^2}$, with $x_j=3j/50$. The contour color transitions from yellow to blue with increasing $j$.}
	\label{fig:df_time_series}
\end{figure*}

In Fig. \ref{fig:df_time_series}, we show three snapshots of $\hat{S}^tf$ for three values of $N$. The initial condition is again $f_0(q,p+\delta p)$ (the same as for Fig. \ref{fig:degenerate_eigenfunctions_subspace_contribution}). Time progresses down the rows, and basis size increases across the columns. Looking first at $N=14$ (rightmost column), the distribution function winds into a one-armed spiral as $t$ increases. Let us compare this to the $N=12$ and $N=10$ columns, focusing on $(2\pi)^{-1}\kappa t = 8$ (second row). Relative to $N=14$, the other two cases appear to have more sophisticated structure, with the complexity increasing as $N$ decreases. This apparent structure is however an artifact of the truncated basis representation in a manner similar to the Gibbs phenomenon. That is, a deficiency of terms in the series over the basis functions leads to a poor representation of the true $f$. 

The basis size set by the polynomial order $N$ determines how well a temporally isolated snapshot of the distribution function can be represented. In a finite dimensional representation, the exact form of $f$ is only known at the initial condition, and is projected onto the basis functions. For $t>0$, every future state $\hat{S}^tf$ is determined from the initial projection according to the eigenvalues of $\hat{A}$. The manner in which the distribution deforms from some $t_1$ to another time $t_2$ depends on the structure of the eigenfunctions and their associated eigenvalues. In the present context, this hinges on the relative spacing of the $\mathcal{O}(\varphi)$ eigenvalues within each degenerate subspace. As discussed in Section \ref{section:eigenvectors_of_A} and demonstrated in Fig. \ref{fig:degenerate_eigenfunctions_subspace_contribution}, the mechanism for spiral formation is the out of phase rotation of the degenerate subspace eigenvectors. The temporal characteristics of this process are determined by the split eigenvalues in Fig. \ref{fig:split_eigenvalues}. This occurs within each degenerate subspace, and the total contribution from the mechanism is encapsulated in the rightmost term in equation \ref{eq:DF_solution_anharmonic}.

 The number of eigenvalues, and their multiplicities in the $\hat{A}_0$ spectrum increase with $N$. For $\lambda_j^{(0)}=-\mathrm{i}\kappa$, we have $d_j=5,6,7$ for $N=10,12,14$. When the degenerate eigenvalues have been split by the anharmonic potential, the multiplicity prescribes the number of distinct rates of rotation for the subspace eigenvectors. This can also be understood as an increase in frequency domain coverage in the sense of a Fourier transform. The length scale set by $N$ determines which structures can be well represented by the basis regardless of $t$. The fanning out in frequency shown in Fig. \ref{fig:split_eigenvalues} coupled with the form of the $\psi_{j,k}$ determines the resolvability of spatiotemporal structure. 
 

\section{Discussion}\label{section:discussion}

The resolvability of structure in $f$ depends on its representation, or means of observation. When one chooses a number of particles in an $N$-body simulation, a grid resolution in an equation solver, or a finite dimensional basis, a scale is imposed. In this work, we aimed to highlight the relationship between imposed length scale from dimension of representation, and multiplicity of eigenvalues in the spectrum of $\hat{A}_0$. Further, that in such a case splitting of the degenerate eigenvalues by $\hat{A}_4$ drives mixing in $f$. 

For the spectrum of $\hat{A}_0$, the number of unique degenerate eigenvalues, and their average multiplicity increases with $M$, the size of the basis. In this construction, spiral formation is achieved through a linear combination of structures that span the entire $(q,p)$ plane (Fig. \ref{fig:degenerate_eigenfunctions_1}). This is in contrast to the picture described in Section \ref{section:introduction}, in which essentially every infinitesimal annulus of phase space volume at a different radius from the origin orbits with a different frequency. In \cite{banik2022}, phase mixing is achieved through a linear response term of the form $\mathrm{e}^{in(\theta-\Omega(I)t)}$, where $(\theta,I)$ are angle-action coordinates, and $\Omega$ is the oscillatory frequency of the orbits. Since the frequencies depends on the actions, orbits at different actions will possess varying frequencies, leading to deformation of $f$. 

One could suppose that for an infinite dimensional basis, $f$ decomposes into an infinite set of delta functions of $q$ and $p$, each nonzero at a different point in the plane. In this case, all of the different packets of phase space density may have different orbital frequencies, and we obtain the original picture of the process, essentially moving to a discrete particle representation. Given the premise of increasing dimension corresponding to increasing degeneracy of the eigenvalues, we suppose that in the limit case of an infinite dimensional representation, the discretely split degenerate eigenvalues may become the continuous bands described in \cite{mathur1990} and \cite{weinberg1991}.

The analysis here omits a self-interaction potential. This means that $\hat{A}$ does not depend on $t$, and the integral in equation \ref{eq:koopman_operator} is trivial. Were this not the case, the operator solution $\hat{U}^t$ would take the form of a Dyson series, obtained by iterative solution of the implicit equation $\hat{U}^t = 1 + \int_0^td\tau\hat{A}(\tau)\hat{U}^\tau$ (see for example, \cite{sakurai2017}). In doing this, it would be sensible to separate the self-interaction contribution to the Hamiltonian, and parameterize its strength with for example, $\alpha$. Computing the necessary Dyson series to some order in $\alpha$  is technically possible. The main constraint is that the matrix element integrals cannot necessarily be evaluated analytically for a non-polynomial potential in our chosen basis. Using the exact one-dimensional Green's function of the Poisson equation poses a challenge in this regard. One option is to adopt a Hamiltonian Mean Field (HMF) approach as in \cite{inagaki1993}, replacing the absolute value function with a polynomial that preserves some desired properties. In this case, the self-interaction can be expressed in terms of the moments of $f$ as in Section \ref{section:moments_A0}, which work nicely with the functional formalism used here. The simplest case would be a quadratic interaction, $v(q,q')\propto(q-q')^2$. In that case, the self-interaction is a quadratic potential well that tracks the expectation value of position with respect to $f$, which is the moment $\hat{U}^tF_{1,0}$.

In this work we considered systems with only a single spatial degree of freedom. This is applicable to the vertical motions of Solar Neighborhood stars in the limit that vertical and in-plane motions decouple. The general approach described in this work may be extended to higher dimensional phase spaces in order to incorporate radial and/or azimuthal motions. The general framework requires that CBE solutions reside in a Hilbert space $\mathscr{H}$, and that there exists a dual space of the corresponding functionals $\mathscr{H}^*$. We can in principle assume this for any number of spatial degrees of freedom. It is when computing a representation of the Koopman generator $\hat{A}$ that we must handle the problem a little differently. First it should be noted that increasing the dimension of the phase space will require additional degrees of freedom on the basis functions. In the case of a truncated finite dimensional basis, this means that in order to achieve a comparable resolution to the calculations here, a larger number of basis functions is necessary.

Assuming a Milky Way-like geometry, we briefly outline here possible choices for the basis functions, and discuss how one might proceed. In Section \ref{section:matrix}, we used Gaussian-Hermite functions for the vertical position and momentum, and can similarly use such functions for the radial and azimuthal momenta. For the radial position, we need orthogonal functions on the interval $[0,\infty)$ that approach zero at infinity. These criteria are met by Laguerre polynomials weighted by a radially decaying exponential function. For the azimuthal position we could use any set of periodic functions orthogonal on $[0,2\pi]$.  

If the increased matrix size poses a computational barrier, one may consider studying the system dynamics with expectation values of particular functions rather than the distribution function itself. One approach is to look at the mean vertical position and/or momentum as a function the radial and azimuthal coordinates as in for example \cite{chequers2017}. These partial moments have been represented in terms of Koopman generator eigenfunctions computed with DMD in \cite{widrow2020conference}. The functional formalism used in this work can be extended to various moments by transforming the Morrison bracket definition of $\hat{A}$ (equation \ref{eq:generator_definition}) into a form prescribed by \cite{kupershmidt1978}. This yields an equation of motion for moments rather than the distribution function itself. A description of this and its correspondence to Jeans' equations can be found in Chapter 3 of \cite{darling_thesis_2024}.



\section{Conclusions}\label{section:conclusion}

We have described a procedure for determining the CBE dynamics of $f$, by mapping the problem into a set of functionals $G[f]$ which satisfy a Schr{\"o}dinger-type equation. To calculate explicit matrix representations of operators, we adopted a finite dimensional set of basis functions that impose a minimum length scale on $f$. For an illustrative example of phase mixing, we treated a quartic potential perturbatively with respect to a harmonic oscillator solution. We observed that the multiplicity of harmonic oscillator eigenvalues increases with the dimension of the finite basis representation. Subsequently, we showed that the anharmonic potential splits the degenerate eigenvalues, leading to the amplitude dependent frequencies characteristic of phase mixing.


In Section \ref{section:eigenvectors_of_A}, we computed the $\mathcal{O}(\varphi)$ corrected eigenfunctions of $\hat{A}$. These structures were segmented into subspaces, with the particular groupings set by the spectrum of $\hat{A}_0$ according to the degeneracy of its eigenvalues. Different classes of structure (for example bending or breathing) were associated with each subspace. The scenario depicted in Fig. \ref{fig:degenerate_eigenfunctions_subspace_contribution} can be understood as the kinematic response to a bending mode perturbation, projected onto the $\lambda_j^{(0)}=-\mathrm{i}\kappa$ subspace. This demonstrated how the standard bending mode structure deforms to contain spiral structure, similar to the observations in \cite{darling2019}. This is of interest in the context of the Gaia spiral, especially with respect to the hypothesis that the observed structure stems from a bending mode excitation \citep{darling2018}. The essential takeaway from this is that the $\hat{A}$ spectrum derived from a solvable model like the harmonic oscillator can be used to model both processes characteristic of self-gravity, and phase mixing. For the approach taken in \cite{darling2019}, one could compute basis functions for the evolution of $f$ which were suited to representing the particular dynamics of the simulation they were derived from. The form of the basis functions derived from DMD was heavily dependent on the parameters of the simulation. They were especially disparate between simulations dominated by self-gravity and anharmonic forcing. In the formalism used here, we suggest that one can construct a single basis suited to both cases. The trade-off needed for this is that instead of a single eigenfunction representative of a principal component of the dynamics, one has a subspace of dimension $d_j\geq1$ (usually $>$). The eigenfunctions spanning a subspace would at first glance appear suited only to represent the highly symmetric oscillatory structures typical of the harmonic oscillator spectrum, but eigenvalue splitting in proportion to anharmonic forcing magnitude allows for out of phase rotation. Such rotation facilitates spiral structure formation in the subspace contributions and consequently in the distribution function.


In modeling the Gaia spiral, it is possible that there does not exist a purely kinematic description which yields the observed structure. Entertaining that possibility, one must ask how self-interaction changes the standard mechanism for the formation of phase space spirals. It is feasible to consider self-interaction and comparatively static anharmonic forcing as competing effects, with their relative magnitudes impacting the spatiotemporal structure of the distribution function. This relative dominance has previously been quantified by a ``live fraction" parameter, denoted $\alpha$ \citep{darling2019,bennet2021,darling2021}. It was observed in \cite{darling2019} that when evolution is primarily driven by self-interaction ($\alpha\simeq0.8$), the dominant contributions to the Koopman spectrum resemble the bending and breathing modes that appear in the harmonic oscillator spectrum (Fig. \ref{fig:shm_eigenfunctions_bending_mode} and \ref{fig:shm_eigenfunctions_breathing_mode}). As the relative strength of the self-interaction was increased, the structures associated with the harmonic oscillator spectrum were deformed to contain spiral structure, resembling quite closely the sum over the $\lambda_j^{(0)}=-\mathrm{i}\kappa$ subspace in Fig. \ref{fig:degenerate_eigenfunctions_subspace_contribution} for $t>0$. We suppose that the spiral-bending and spiral-breathing contributions to the Koopman spectrum observed numerically in \cite{darling2019} can be modeled by the degenerate subspaces of the perturbed harmonic oscillator spectrum associated with their respective structures. For the quartic Hamiltonian, the eigenvalue splitting mechanism represents the effect of the anharmonic forcing. The self-interaction can be treated with time-dependent perturbation theory. Specifically this involves the more general Dyson series for $\frac{\partial \hat{A}}{\partial t}\neq 0$ discussed in Section \ref{section:discussion}. This treatment can be applied to each subspace individually, so one can observe the effect of self-interaction on the spiral-bending and spiral-breathing modes in a similar way to \cite{darling2019}, but with an analytic model.

A crucial assumption here is that one can determine a basis sufficient for the evolution of the distribution function. If that is the case, the problem is reduced to determining the set of coefficients that produce the correct linear combination. It is the relative magnitudes of the coefficients that determine the form of $f$ at any particular time, but it is the relative phase of the time-dependent coefficients that prescribe the evolution and mixing. Acknowledging the possibility that there is no fully time-independent mapping from the hypothetically ``correct" initial condition to the observed structure, the effect of time-dependent forces from the self-interaction of the distribution is essential for successful modeling. We suggest that a subspace-wise analysis of the interplay between anharmonic forcing and self-interaction is a preferable way forward. Preliminary calculations using the time-dependent $\hat{A}$ Dyson series indicate that addition of a self-interaction term to the quartic Hamiltonian leads to a time-varying rate of mixing. This is corroborated by fully self-consistent \citep{darling2018} and live fraction based \citep{darling2019} numerical experiments. In all cases, it was observed that a self-interacting distribution undergoes mostly kinematic phase mixing at early times to form a spiral, but the winding slows down after the initial formation. A detailed exploration of this is beyond the scope of the present work, and is left to a future article.




It is also of interest to further investigate the scale-dependent mixing process studied here in the context of other work around coarse-grained evolution of the CBE, as in \cite{chavanis1996} or \cite{chavanis2005}. In the latter, one of the definitions of a coarse-grained $f$ is a windowed functional. That is, $f$ is taken in convolution with some kernel that sets the representation scale. Preliminary work suggests that the equation of motion in that case is equivalent to equation \ref{eq:koopman_ode} up to a diffusion current term. 

\section*{Acknowledgements}

This work was supported by a Discovery Grant with the Natural Sciences and Engineering Research Council of Canada. We are thankful to Mike Petersen for insightful comments. We also thank Scott Tremaine, Francesco Cellarosi, Aaron Vincent, and Stephen Hughes for helpful discussions.  

\section*{Data Availability}

Calculations were carried out in MATLAB. Custom software used for this paper is available upon reasonable request. Colormaps used in contour plots are thanks to \cite{cmocean2016}.



\bibliographystyle{mnras}
\bibliography{thesis_references} 




\appendix

\section{Preliminary Definitions}\label{A:definitions}

\subsection{Inner products}

The inner product in $\mathscr{H}$ is

\begin{equation}\label{eq:inner_product}
	\langle g_1,g_2\rangle = \int_\mathcal{D}dqdp \ g_1(q,p)g_2^\dagger(q,p),
\end{equation}

\noindent where $^\dagger$ indicates complex conjugation. This product possesses conjugate symmetry, $\langle g_1,g_2\rangle^\dagger = \langle g_2,g_1\rangle$. For a set of linearly independent basis functions $\{e_{j,k}\}$ as defined in Section \ref{section:setup}, the inner product facilitates projection of an arbitrary function $g\in\mathscr{H}$. Such an expansion may be expressed as

\begin{equation}\label{eq:function_space_expansion}
	g(q,p) = \sum_{j=0}^{\infty}\sum_{k=0}^{\infty}\langle e_{j,k},g\rangle e_{j,k}(q,p).
\end{equation}

The inner product in $\mathscr{H}^*$ is

\begin{equation}\label{eq:dual_inner_product}
	\langle G_1,G_2\rangle_* = \bigg\langle \frac{\delta G_2}{\delta f}, \frac{\delta G_1}{\delta f}\bigg\rangle. 
\end{equation}

\noindent An expansion analogous to equation \ref{eq:function_space_expansion} applies to arbitrary functionals $G\in\mathscr{H}^*$.

\subsection{Hermite polynomials}

The Hermite polynomials $P_j(x)$ introduced in Section \ref{section:choice_of_basis} are defined by the Rodrigues formula,

\begin{equation}
	P_j(x) = (-1)^n\mathrm{e}^{x^2}\frac{d^j}{dx^j}\left(\mathrm{e}^{-x^2}\right).
\end{equation}

\noindent Hermite polynomials satisfy the following two recurrence relations:

\begin{equation}\label{eq:hermite_property_1}
	\frac{d}{dx}P_j(x) = 2jP_{j-1}(x),
\end{equation}

\begin{equation}\label{eq:hermite_property_2}
	xP_j(x) = \frac{1}{2}P_{j+1}(x) + jP_{j-1}(x).
\end{equation} 

The normalization constants for the Gaussian-Hermite basis functions defined in Section \ref{section:choice_of_basis} are

\begin{equation}
	N_{j,k} = \frac{Z}{\sqrt{\pi 2^{j+k} j!k!}}.
\end{equation}

Our chosen bivariate functions inherit orthogonality from the Hermite polynomials. Explicitly,

\begin{equation}\label{eq:basis_orthogonality}
	\langle e_{j,k},e_{j',k'}\rangle = \delta_{j'}^{j}\delta_{k'}^k.
\end{equation}

\noindent This applies to the entire real number line in both variables, so our domain $\mathcal{D}$ is set to the entire phase space.

\section{Perturbative treatment of degenerate eigenvalues}\label{A:degenerate_perturbation}

We aim to find eigenvectors $\boldsymbol{\psi}_{j,k}(\varphi)$ and eigenvalues $\lambda_{j,k}(\varphi)$ that satisfy the perturbed eigenvalue problem 

\begin{equation}\label{eq:perturbed_eigen_problem}
	\left(\bm{\mathsf{A}}_\mathsf{0}+ \varphi \bm{\mathsf{A}}_\mathsf{4} \right)\boldsymbol{\psi}_{j,k} = \lambda_{j,k}\boldsymbol{\psi}_{j,k}.
\end{equation}

\noindent This is achieved by assuming a power series in $\varphi$ for both quantities,

\begin{equation}\label{eq:perturbation_power_series}
	\begin{aligned}
		\lambda_{j,k}(\varphi) &= \lambda_{j,k}^{(0)} + \varphi\lambda_{j,k}^{(1)} + \mathcal{O}(\varphi^2), \\
		\boldsymbol{\psi}_{j,k}(\varphi) &= \boldsymbol{\psi}_{j,k}^{(0)} + \varphi\boldsymbol{\psi}_{j,k}^{(1)} + \mathcal{O}(\varphi^2).
	\end{aligned}
\end{equation}

\noindent This requires that both the eigenvectors and eigenvalues vary smoothly with respect to change in the perturbation parameter $\varphi$. Since the eigenvalues are scalar, this is the case by default. We assure a smooth variation in the eigenvectors by the procedure described in Section \ref{section:simultaneous_eigenvectors}. 

Substituting the assumed power series into equation \ref{eq:perturbed_eigen_problem}, and equating terms of equal power in $\varphi$, one finds a set of equations. Those corresponding to the first three powers of $\varphi$ are, 

\begin{equation}\label{eq:degenerate_PT_equal_powers}
	\begin{aligned}
		\left(\bm{\mathsf{A}}_0 - \lambda_{j}^{(0)}\right)\boldsymbol{\psi}_{j,k}^{(0)} &= 0, \\
		\left(\bm{\mathsf{A}}_0 - \lambda_{j}^{(0)}\right)\boldsymbol{\psi}_{j,k}^{(1)} &= \left(\lambda^{(1)}_{j,k} - \bm{\mathsf{A}}_4\right)\boldsymbol{\psi}_{j,k}^{(0)}, \\
		\left(\bm{\mathsf{A}}_0 - \lambda_j^{(0)}\right)\boldsymbol{\psi}^{(2)}_{j,k} &= \left(\lambda_{j,k}^{(1)} - \bm{\mathsf{A}}_4\right)\boldsymbol{\psi}_{j,k}^{(1)} + \lambda_{j,k}^{(2)}\boldsymbol{\psi}_{j,k}^{(0)}.
	\end{aligned}
\end{equation}

\subsection{Degenerate case: eigenvalues}\label{A:degenerate_eigenvalues}

To determine the first order correction to the eigenvalues, we begin by contracting the $\mathcal{O}(\varphi)$ case (second line) in equation \ref{eq:degenerate_PT_equal_powers} with ${\boldsymbol{\psi}^{(0)}_{j,l}}^\dagger$. That is

\begin{equation}
	{\boldsymbol{\psi}^{(0)}_{j,l}}^\dagger\cdot
	\left(\bm{\mathsf{A}}_0 - \lambda_{j}^{(0)}\right)\boldsymbol{\psi}_{j,k}^{(1)} = {\boldsymbol{\psi}^{(0)}_{j,l}}^\dagger\cdot\left(\lambda^{(1)}_{j,k} - \bm{\mathsf{A}}_4\right)\boldsymbol{\psi}_{j,k}^{(0)}.
\end{equation}

\noindent Since ${\boldsymbol{\psi}^{(0)}_{j,k}}^\dagger$ is a left-eigenvector of $\bm{\mathsf{A}}_0$, ${\boldsymbol{\psi}^{(0)}_{j,k}}^\dagger\bm{\mathsf{A}}_0=\lambda_{j,k}^{(0)}{\boldsymbol{\psi}^{(0)}_{j,k}}^\dagger$, and the left hand side is zero. We are left with 

\begin{equation}\label{eq:degenerate_eigenvalues_1}
	\lambda_{j,k}^{(1)}{\boldsymbol{\psi}^{(0)}_{j,l}}^\dagger \cdot \boldsymbol{\psi}_{j,k}^{(0)} ={\boldsymbol{\psi}^{(0)}_{j,l}}^\dagger\cdot \bm{\mathsf{A}}_4\boldsymbol{\psi}_{j,k}^{(0)}.	
\end{equation}

\noindent Since ${\boldsymbol{\psi}^{(0)}_{j,l}}^\dagger\cdot \boldsymbol{\psi}_{j,k}^{(0)} = \delta_l^k$, equation \ref{eq:degenerate_eigenvalues_1} then reduces to 

\begin{equation}\label{eq:degenerate_eigenvalues_correction_1}
	\lambda_{j,k}^{(1)} = {\boldsymbol{\psi}^{(0)}_{j,k}}^\dagger \cdot \bm{\mathsf{A}}_4\boldsymbol{\psi}_{j,k}^{(0)}.
\end{equation}

\noindent That is, the $\mathcal{O}(\varphi)$ corrections to the degenerate eigenvalues are the diagonal entries of the perturbation, $\bm{\mathsf{A}}_4$, when it is projected onto the degenerate subspace corresponding to $\lambda^{(0)}_{j,k}$. In other words, if we project $\bm{\mathsf{A}}_4$ onto a degenerate subspace, the eigenvalues of the projected matrix are the first order corrections $\lambda_{j,k}^{(1)}$. The first order correction $\lambda_{j,k}^{(1)}$ in equation \ref{eq:degenerate_eigenvalues_correction_1} goes into the power series definition for the eigenvalues of $\hat{A}$ in Section \ref{section:eigenvalues_of_A} (equation \ref{eq:degenerate_values_series}).

\subsection{Degenerate case: eigenvectors}\label{A:degenerate_eigenvectors}

In general, the eigenvectors of $\bm{\mathsf{A}}$ may contain components from the entire space, including vectors from both in and not in the degenerate subspace. That is, we want to know how to  project $\boldsymbol{\psi}_{j,k}^{(1)}$ onto the sets of $\boldsymbol{\psi}_{j}^{(0)}$ and $\boldsymbol{\psi}_{j,k}^{(0)}$. It follows that we can do this if we have general expressions for both of the contractions, $ {\boldsymbol{\psi}_{j}^{(0)}}^\dagger\cdot \boldsymbol{\psi}_{j,k}^{(1)} $ and $ {\boldsymbol{\psi}_{j,k}^{(0)}}^\dagger\cdot \boldsymbol{\psi}_{j,k}^{(1)} $.

We will treat both cases separately, first computing the contribution from the subset of the complete basis that excludes the degenerate subspace. We begin by contracting the first order case in equation \ref{eq:degenerate_PT_equal_powers} with ${\boldsymbol{\psi}_{j}^{(0)}}^\dagger$. We have

\begin{equation}\label{eq:degenerate_eigenvectors_1}
	{\boldsymbol{\psi}_{l}^{(0)}}^\dagger\cdot 
	\left(\bm{\mathsf{A}}_0 - \lambda_{j}^{(0)}\right)\boldsymbol{\psi}_{j,k}^{(1)} = {\boldsymbol{\psi}_{l}^{(0)}}^\dagger\cdot \left(\lambda^{(1)}_{j,k} - \bm{\mathsf{A}}_4\right)\boldsymbol{\psi}_{j,k}^{(0)}. 
\end{equation}

\noindent Similar to when we computed the eigenvalue correction, we note that ${\boldsymbol{\psi}_{j}^{(0)}}^\dagger\bm{\mathsf{A}}_0 = \lambda_{j}^{(0)}{\boldsymbol{\psi}_{j}^{(0)}}^\dagger  $. Additionally, we know that ${\boldsymbol{\psi}_{l}^{(0)}}^\dagger$ and $\boldsymbol{\psi}_{j,k}^{(0)}  $ are orthogonal for all $ l,j,k $, since one of them is within the degenerate subspace and the other is not. From these two arguments, we can simplify equation \ref{eq:degenerate_eigenvectors_1} to 

\begin{equation}\label{eq:degenerate_eigenvectors_2}
	{\boldsymbol{\psi}_{l}^{(0)}}^\dagger\cdot \boldsymbol{\psi}_{j,k}^{(1)} = \frac{{\boldsymbol{\psi}_{l}^{(0)}}^\dagger\cdot \bm{\mathsf{A}}_4\boldsymbol{\psi}_{j,k}^{(0)}}{\lambda_{l}^{(0)} - \lambda_{j}^{(0)}}.
\end{equation}

For the degenerate subset contribution, we do not get any new information by attempting a contraction of $ {\boldsymbol{\psi}_{j,k}^{(0)}}^\dagger $ with the first order case in equation \ref{eq:degenerate_PT_equal_powers}. Let us instead try the $\mathcal{O}(\varphi^2)$ equation. We take 

\begin{equation}
\begin{aligned}
	{\boldsymbol{\psi}_{j,l}^{(0)}}^\dagger \cdot \left(\bm{\mathsf{A}}_0 - \lambda_j^{(0)}\right)\boldsymbol{\psi}^{(2)}_{j,k} =  {\boldsymbol{\psi}_{j,l}^{(0)}}^\dagger \cdot \left(\lambda_{j,k}^{(1)} - \bm{\mathsf{A}}_4\right)\boldsymbol{\psi}_{j,k}^{(1)} \\
+ \lambda_{j,k}^{(2)} {\boldsymbol{\psi}_{j,l}^{(0)}}^\dagger \cdot \boldsymbol{\psi}_{j,k}^{(0)}.
\end{aligned}
\end{equation}

\noindent Note that we use the same index $j$ on the first index for the introduced vector, as if we allowed for a different index, we would be considering the case that one degenerate subspace is projected onto another. All of the distinct subspaces are orthogonal, so this can only result in zero. 

We can proceed by again noting that ${\boldsymbol{\psi}_{j,l}^{(0)}}^\dagger\bm{\mathsf{A}}_0=\lambda_{j,l}^{(1)}$. Since $\lambda_{j,l} = \lambda_{j} \ \forall \ l$, the left hand side is zero. Further, ${\boldsymbol{\psi}_{j,l}^{(0)}}^\dagger \cdot \boldsymbol{\psi}_{j,k}^{(0)} = \delta_l^k$, so we are are left with 

\begin{equation}\label{eq:degenerate_eigenvectors_3}
	\lambda_{j,k}^{(2)} \delta_l^k =  -{\boldsymbol{\psi}_{j,l}^{(0)}}^\dagger \cdot \left(\lambda_{j,k}^{(1)} - \bm{\mathsf{A}}_4\right)\boldsymbol{\psi}_{j,k}^{(1)}.
\end{equation}

\noindent Of course, we do not know $ \boldsymbol{\psi}_{j,k}^{(1)} $, as that is what we are trying to determine here. We do know however that it can have contributions from $ \boldsymbol{\psi}_j^{(0)} $ and  $\boldsymbol{\psi}_{j,l}^{(0)}$. Let us suppose that it takes the form 

\begin{equation}\label{eq:degenerate_eigenvectors_4}
	\boldsymbol{\psi}_{j,k}^{(1)} = \sum_{m\neq j }^M \left({\boldsymbol{\psi}_{m}^{(0)}}^\dagger \cdot \boldsymbol{\psi}_{j,k}^{(1)}\right)\boldsymbol{\psi}_m^{(0)} + \sum_{l}^{d_j} \left({\boldsymbol{\psi}_{j,l}^{(0)}}^\dagger\cdot \boldsymbol{\psi}_{j,k}^{(1)}\right) \boldsymbol{\psi}_{j,l}^{(0)}.
\end{equation}

\noindent We have already found what we need to express the first sum explicitly in equation \ref{eq:degenerate_eigenvectors_2}. With this, we may write 

\begin{equation}
	\boldsymbol{\psi}_{j,k}^{(1)} = \sum_{m\neq j }^M \frac{{\boldsymbol{\psi}_{m}^{(0)}}^\dagger\cdot  \bm{\mathsf{A}}_4\boldsymbol{\psi}_{j,k}^{(0)}}{\lambda_{m}^{(0)} - \lambda_{j}^{(0)}}
	\boldsymbol{\psi}_m^{(0)} + \sum_{l}^{d_j} \left({\boldsymbol{\psi}_{j,l}^{(0)}}^\dagger\cdot \boldsymbol{\psi}_{j,k}^{(1)}\right) \boldsymbol{\psi}_{j,l}^{(0)}.
\end{equation}

\noindent Continuing with equation \ref{eq:degenerate_eigenvectors_3}, we have 

\begin{equation}
	\begin{aligned}
		\lambda_{j,k}^{(2)} \delta_l^k =  &-\lambda_{j,k}^{(1)} \sum_{m\neq j }^M \frac{{\boldsymbol{\psi}_{m}^{(0)}}^\dagger\cdot  \bm{\mathsf{A}}_4\boldsymbol{\psi}_{j,k}^{(0)}}{\lambda_{m}^{(0)} - \lambda_{j}^{(0)}}
		{\boldsymbol{\psi}_{j,l}^{(0)}}^\dagger \cdot \boldsymbol{\psi}_m^{(0)} \\
		&+
		\sum_{m\neq j }^M \frac{{\boldsymbol{\psi}_{m}^{(0)}}^\dagger \cdot \bm{\mathsf{A}}_4\boldsymbol{\psi}_{j,k}^{(0)}}{\lambda_{m}^{(0)} - \lambda_{j}^{(0)}} {\boldsymbol{\psi}_{j,l}^{(0)}}^\dagger\cdot \bm{\mathsf{A}}_4 
		\boldsymbol{\psi}_m^{(0)}\\
		&-\sum_{n}^{d_j} \lambda_{j,k}^{(1)} \left({\boldsymbol{\psi}_{j,n}^{(0)}}^\dagger \cdot \boldsymbol{\psi}_{j,k}^{(1)}\right)
		\left({\boldsymbol{\psi}_{j,l}^{(0)}}^\dagger \cdot 
		\boldsymbol{\psi}_{j,n}^{(0)}\right) \\
		&+
		\sum_{n}^{d_j} \left({\boldsymbol{\psi}_{j,n}^{(0)}}^\dagger \cdot \boldsymbol{\psi}_{j,k}^{(1)}\right)
		{\boldsymbol{\psi}_{j,l}^{(0)}}^\dagger \cdot \bm{\mathsf{A}}_4
		\boldsymbol{\psi}_{j,n}^{(0)}
	\end{aligned}
\end{equation}

\noindent The first term on the right hand side is zero since $ {\boldsymbol{\psi}_{j,l}^{(0)}}^\dagger \cdot \boldsymbol{\psi}_m^{(0)}  $ are orthogonal for $m\neq j $. The sum in the term on the right hand side collapses because $ {\boldsymbol{\psi}_{j,l}^{(0)}}^\dagger \cdot \boldsymbol{\psi}_{j,n}^{(0)} = \delta_l^n $. Making these simplifications leaves us with, 

\begin{equation}
\begin{aligned}
	\lambda_{j,k}^{(2)} \delta_l^k =  
&\sum_{m\neq j }^M \frac{{\boldsymbol{\psi}_{m}^{(0)}}^\dagger \cdot  \bm{\mathsf{A}}_4\boldsymbol{\psi}_{j,k}^{(0)}}{\lambda_{m}^{(0)} 
	- \lambda_{j}^{(0)}} {\boldsymbol{\psi}_{j,l}^{(0)}}^\dagger  \cdot \bm{\mathsf{A}}_4 
\boldsymbol{\psi}_m^{(0)}\\
&-\lambda_{j,k}^{(1)} \left({\boldsymbol{\psi}_{j,l}^{(0)}}^\dagger  \cdot \boldsymbol{\psi}_{j,k}^{(1)}\right)
\\
&+
\sum_{n}^{d_j} \left({\boldsymbol{\psi}_{j,n}^{(0)}}^\dagger \cdot \boldsymbol{\psi}_{j,k}^{(1)} \right)
{\boldsymbol{\psi}_{j,l}^{(0)}}^\dagger \cdot  \bm{\mathsf{A}}_4
\boldsymbol{\psi}_{j,n}^{(0)}.
\end{aligned}
\end{equation}

\noindent Since the degenerate subspace basis vectors $\boldsymbol{\psi}_{j,k}^{(0)}$ are chosen such that they diagonalize $\bm{\mathsf{A}}_4$, the only nonzero matrix elements $ {\boldsymbol{\psi}_{j,l}^{(0)}}^\dagger \cdot \bm{\mathsf{A}}_4
\boldsymbol{\psi}_{j,n}^{(0)} $ are those along the diagonal. We may therefore simplify this term as  

\begin{equation}
\begin{aligned}
	\lambda_{j,k}^{(2)} \delta_l^k =  
&\sum_{m\neq j }^M \frac{{\boldsymbol{\psi}_{m}^{(0)}}^\dagger \cdot \bm{\mathsf{A}}_4\boldsymbol{\psi}_{j,k}^{(0)}}{\lambda_{m}^{(0)} - \lambda_{j}^{(0)}} {\boldsymbol{\psi}_{j,l}^{(0)}}^\dagger  \cdot \bm{\mathsf{A}}_4 
\boldsymbol{\psi}_m^{(0)} \\
&-\lambda_{j,k}^{(1)} \left({\boldsymbol{\psi}_{j,l}^{(0)}}^\dagger \cdot \boldsymbol{\psi}_{j,k}^{(1)}\right)
+
\sum_{n}  \lambda_{j,l}^{(1)}\delta_l^n  \left( {\boldsymbol{\psi}_{j,n}^{(0)}}^\dagger \cdot \boldsymbol{\psi}_{j,k}^{(1)} \right)
.
\end{aligned}
\end{equation}

\noindent Collapsing the sum according to the $\delta_l^n$ we are left with 

\begin{equation}
\begin{aligned}
	\lambda_{j,k}^{(2)} \delta_l^k =  
\sum_{m\neq j }^M \frac{{\boldsymbol{\psi}_{m}^{(0)}}^\dagger \cdot  \bm{\mathsf{A}}_4\boldsymbol{\psi}_{j,k}^{(0)}}{\lambda_{m}^{(0)} - \lambda_{j}^{(0)}} {\boldsymbol{\psi}_{j,l}^{(0)}}^\dagger \cdot \bm{\mathsf{A}}_4 
\boldsymbol{\psi}_m^{(0)}\\
+	\left(\lambda_{j,l}^{(1)} - \lambda_{j,k}^{(1)}\right){\boldsymbol{\psi}_{j,l}^{(0)}}^\dagger \cdot \boldsymbol{\psi}_{j,k}^{(1)}.
\end{aligned}
\end{equation}

\noindent This equation has two cases. First, if $l\neq k$, we obtain the explicit expression for ${\boldsymbol{\psi}_{j,l}^{(0)}}^\dagger \cdot \boldsymbol{\psi}_{j,k}^{(1)}  $ that we have been looking for. That is

\begin{equation}
	{\boldsymbol{\psi}_{j,l}^{(0)}}^\dagger \cdot \boldsymbol{\psi}_{j,k}^{(1)} = 
	\frac{1}{\lambda_{j,l}^{(1)} - \lambda_{j,k}^{(1)}}\sum_{m\neq j }^M \frac{{\boldsymbol{\psi}_{m}^{(0)}}^\dagger \cdot  \bm{\mathsf{A}}_4\boldsymbol{\psi}_{j,k}^{(0)}}{\lambda_{m}^{(0)} - \lambda_{j}^{(0)}} {\boldsymbol{\psi}_{j,l}^{(0)}}^\dagger \cdot \bm{\mathsf{A}}_4 
	\boldsymbol{\psi}_m^{(0)}.
\end{equation}

\noindent Taking the case $l=k$ yields the $\mathcal{O}(\varphi^2)$ correction to the eigenvalues. That is,

\begin{equation}\label{eq:second_order_eigenvalues}
	\begin{aligned}
		\lambda_{j,k}^{(2)} =  
		\sum_{m\neq j }^M \frac{{\boldsymbol{\psi}_{m}^{(0)}}^\dagger \cdot  \bm{\mathsf{A}}_4\boldsymbol{\psi}_{j,k}^{(0)}}{\lambda_{m}^{(0)} - \lambda_{j}^{(0)}} {\boldsymbol{\psi}_{j,k}^{(0)}}^\dagger \cdot \bm{\mathsf{A}}_4 
		\boldsymbol{\psi}_m^{(0)}.
	\end{aligned}
\end{equation} 

\noindent These second order corrections are shown in the dotted lines in Fig. \ref{fig:split_eigenvalues}.

Now that we have explicit expressions for both $ {\boldsymbol{\psi}_{m}^{(0)}}^\dagger \cdot \boldsymbol{\psi}_{j,k}^{(1)} $ and $  {\boldsymbol{\psi}_{j,l}^{(0)}}^\dagger \cdot \boldsymbol{\psi}_{j,k}^{(1)}$ we can express the first order corrections to the eigenvectors of $\bm{\mathsf{A}}_4$ as we had outlined in equation \ref{eq:degenerate_eigenvectors_4}.

\begin{equation}
\begin{aligned}
	\boldsymbol{\psi}_{j,k}^{(1)} = &-\sum_{m\neq j }^M
\frac{{\boldsymbol{\psi}_{m}^{(0)}}^\dagger \cdot \bm{\mathsf{A}}_4\boldsymbol{\psi}_{j,k}^{(0)}}{\lambda_{m}^{(0)} - \lambda_{j}^{(0)}}
\boldsymbol{\psi}_m^{(0)} \\
&+ 
\sum_{l\neq k }^{d_j}
\frac{1}{\lambda_{j,l}^{(1)} - \lambda_{j,k}^{(1)}}\sum_{m\neq j }^M \frac{{\boldsymbol{\psi}_{m}^{(0)}}^\dagger \cdot  \bm{\mathsf{A}}_4\boldsymbol{\psi}_{j,k}^{(0)}}{\lambda_{m}^{(0)} - \lambda_{j}^{(0)}} \left( {\boldsymbol{\psi}_{j,l}^{(0)}}^\dagger \cdot \bm{\mathsf{A}}_4 
\boldsymbol{\psi}_m^{(0)} \right)
\boldsymbol{\psi}_{j,l}^{(0)}.
\end{aligned}
\end{equation}

\noindent Factoring, this can be written more compactly as equation \ref{eq:degenerate_eigenvectors_correction_1}, which is the final result we use in Section \ref{section:eigenvectors_of_A}.


\bsp	
\label{lastpage}
\end{document}